\documentclass[a4paper,
twocolumn,
10pt,
unpublished]{quantumarticle}
\pdfoutput=1
\usepackage[utf8]{inputenc}
\usepackage[english]{babel}
\usepackage[T1]{fontenc}
\usepackage{amsmath}
\usepackage{hyperref}
\usepackage{amsthm}
\usepackage{amsfonts}
\usepackage{braket}
\usepackage{enumerate}
\usepackage{graphicx}
\RequirePackage[numbers,sort&compress]{natbib}
\usepackage{subcaption}
\usepackage[dvipsnames]{xcolor}
\hypersetup{
    colorlinks=true,
    linkcolor={teal}, 
    citecolor=Maroon,     %
    filecolor=blue,     %
    urlcolor=Maroon,      %
}

\usepackage[noend,linesnumbered, ruled, vlined]{algorithm2e}
\usepackage{tikz}
\usepackage{lucas}
\usepackage{multirow}

\newcommand{\secref}[1]{Section~\hyperref[#1]{\ref*{#1}}}
\newcommand{\appref}[1]{Appendix~\hyperref[#1]{\ref*{#1}}}
\newcommand{\tabref}[1]{Table~\hyperref[#1]{\ref*{#1}}}
\newcommand{\figref}[1]{Fig.~\hyperref[#1]{\ref*{#1}}}
\newcommand{\sfigref}[2]{Fig.~\hyperref[#1]{\ref*{#1}(#2)}}

\newcommand{\Cj}{C_{\ell}} %
\newcommand{\Cjindex}{\ell} %
\newcommand{\yesrows}{\bullet} %
\newcommand{\norows}{\bot}
\newcommand{\invalid}{\mathbb{I}}
\newcommand{\valid}{\mathbb{V}}
\renewcommand{\vec}[1]{\mathbf{#1}}

\SetKwInput{KwInput}{Input}

\begin{document}
{
\title{Localized statistics decoding for quantum low-density parity-check codes}
}
\author{Timo Hillmann\textsuperscript{\includegraphics[scale=0.01]{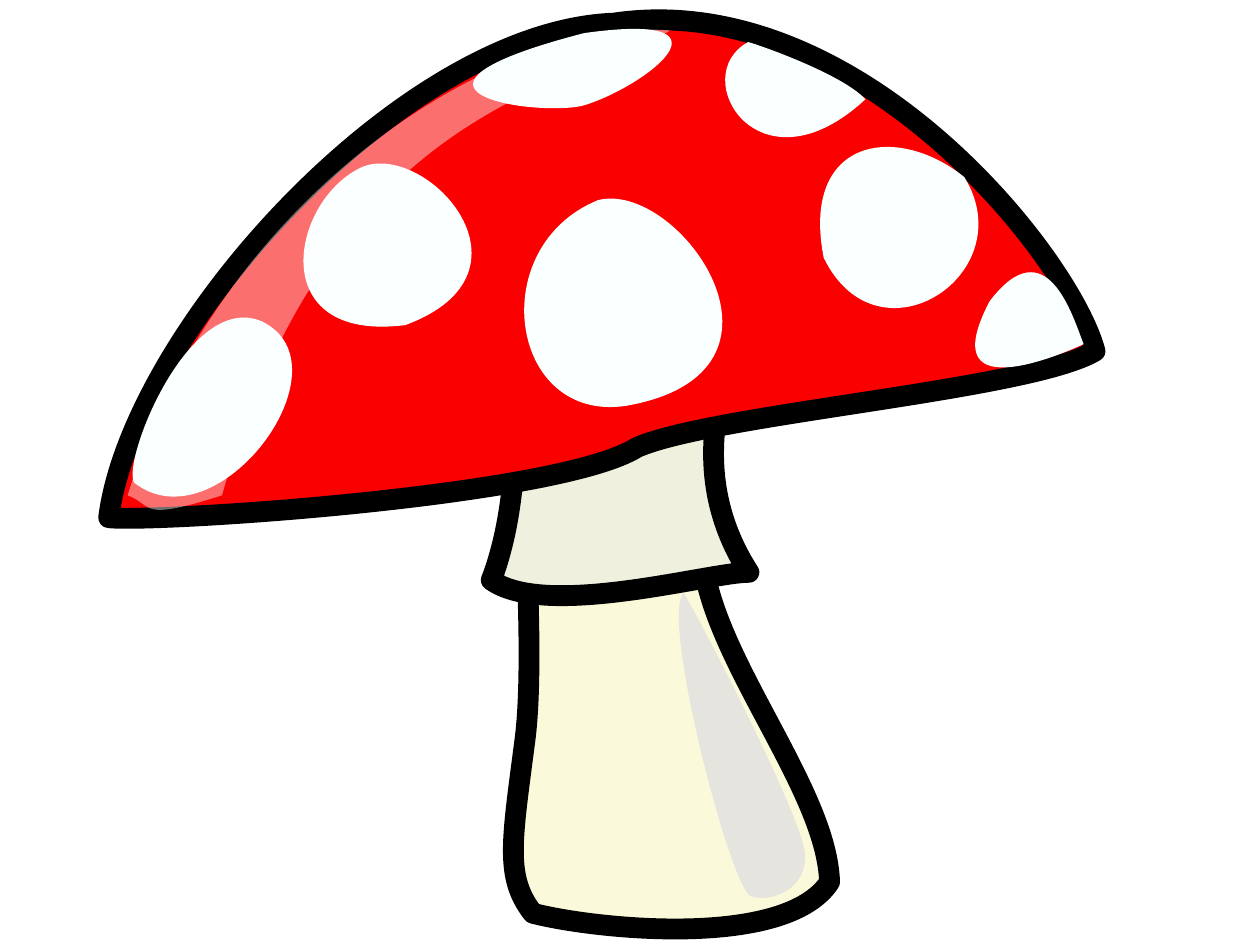}}} %
\affiliation{Chalmers University of Technology, Gothenburg, Sweden}
\email{timo.hillmann@rwth-aachen.de}
\orcid{0000-0002-1476-0647}
\author{Lucas Berent\textsuperscript{\includegraphics[scale=0.01]{figures/Mushroom.pdf}}} %
\affiliation{Technical University of Munich, Germany}
\email{lucas.berent@tum.de}
\orcid{0000-0002-2973-1689}

\author{Armanda O. Quintavalle}
\affiliation{Freie Universit\"at Berlin, Germany}
\orcid{0000-0001-5101-5673}

\author{Jens Eisert}
\affiliation{Freie Universit\"at Berlin, Germany}
\affiliation{Helmholtz-Zentrum Berlin f\"ur Materialien und Energie, Germany}
\orcid{0000-0003-3033-1292}

\author{Robert Wille}
\affiliation{Technical University of Munich, Germany}
\affiliation{Software Competence Center Hagenberg, Austria}
\orcid{0000-0002-4993-7860}

\author{Joschka Roffe}
\affiliation{Freie Universit\"at Berlin, Germany}
\affiliation{University of Edinburgh, United Kingdom}
\email{joschka@roffe.eu}
\orcid{0000-0001-9202-1156}

\thanks{\\ \includegraphics[scale=0.01]{figures/Mushroom.pdf}Timo Hillmann and Lucas Berent contributed equally.}

\maketitle

\begin{abstract}
Quantum low-density parity-check codes are a promising candidate for fault-tolerant quantum computing with considerably reduced overhead compared to the surface code. 
However, the lack of a practical decoding algorithm remains a barrier to their implementation. 
In this work, we introduce localized statistics decoding, a reliability-guided inversion decoder that is highly parallelizable and applicable to arbitrary quantum low-density parity-check codes. 
Our approach employs a parallel matrix factorization strategy, which we call \emph{on-the-fly elimination}, to identify, validate, and solve local decoding regions on the decoding graph.
Through numerical simulations, we show that localized statistics decoding matches the performance of state-of-the-art decoders while reducing the runtime complexity for operation in the sub-threshold regime. 
Importantly, our decoder is more amenable to implementation on specialized hardware, positioning it as a promising candidate for decoding real-time syndromes from experiments.
\end{abstract}

\section{Introduction}
\emph{Quantum low-density parity-check} (QLDPC) codes~\cite{breuckmann2021quantum} are a promising alternative to the surface code~\cite{dennis_topological_2002, fowler_surface_2012, kitaev_fault-tolerant_2003}. 
Based on established methods underpinning classical technologies such as Ethernet and 5G~\cite{gallager_low-density_1962,richardson_design_2018}, QLDPC codes promise a low-overhead route to fault tolerance~\cite{breuckmann_balanced_2021, panteleev_asymptotically_2022, gottesman_fault-tolerant_2014, leverrier_quantum_2022, panteleev_degenerate_2021, bravyi_high-threshold_2024, xu_constant-overhead_2024}, encoding multiple qubits per logical block as opposed to a single one for the surface code.
While, as a trade-off, QLDPC codes require long-range interactions that can be difficult to implement physically, various architectures allow for those requirements~\cite{simmons_scalable_2024, bluvstein_logical_2023, tzitrin_fault-tolerant_2021, bartolucci_fusion-based_2023}. 
In particular, recent work targeting quantum processors based on neutral atom arrays~\cite{xu_constant-overhead_2024} as well a bi-layer superconducting qubit chip architecture~\cite{bravyi_high-threshold_2024} suggest that QLDPC codes can achieve an order-of-magnitude reduction in overhead relative to the surface code on near-term hardware.

In a quantum error correction circuit, errors are detected by measuring stabilizers yielding a stream of syndrome information. 
The \textit{decoder} is the classical co-processor tasked with performing real-time inference on the measured error syndromes to determine a correction operation that must take place within a time frame less than the decoherence time of the physical qubits.
Full-scale quantum computers will impose significant demands on their decoders, with estimates suggesting that terabytes of decoding bandwidth will be required for real-time processing of syndrome data~\cite{bacon_software_2022, beverland_assessing_2022}. 
As such, decoding algorithms must be as efficient as possible and, in particular, suitable for parallel implementation on specialized hardware~\cite{skoric_parallel_2023}. 

The current gold standard for decoding general QLDPC codes is the \emph{belief propagation plus ordered statistics decoder} (BP+OSD)~\cite{panteleev_degenerate_2021,roffe_decoding_2020}. 
The core of this decoder is the iterative \emph{belief propagation} (BP) algorithm~\cite{mackay_near_1997} that finds widespread application in classical error correction.
Unfortunately, BP decoders are not effective out of the box for QLDPC codes. 
The reason for this shortcoming are so-called \emph{degenerate errors}, that is, physically different errors that are equivalent up to stabilizers and prevent BP from converging~\cite{raveendran_trapping_2021,morris_analysis_2023, panteleev_degenerate_2021}. 
The BP+OSD algorithm augments BP with a post-processing routine based on 
\emph{ordered statistics decoding} (OSD)~\cite{fossorier_soft-decision_1995,fossorier_reliability-based_1998, roffe_decoding_2020, panteleev_degenerate_2021}. 
OSD is invoked if the BP algorithm fails to converge and computes a solution by inverting a full-rank submatrix of the parity check matrix. A specific strength of the BP+OSD decoder lies in its versatility: it achieves good decoding performance across the landscape of quantum 
LDPC codes~\cite{roffe_decoding_2020}.

A significant limitation of the BP+OSD decoder is its large runtime overhead.
This inefficiency stems primarily from the OSD algorithm's inversion step, which relies on Gaussian elimination and has cubic worst-case time complexity in the size of the corresponding check matrix. 
In practice, this is a particularly acute problem, as decoders must be run on large circuit-level decoding graphs that account for errors occurring at any location in the syndrome extraction circuit. 
This shortcoming constitutes a known barrier to the experimental implementation of efficient quantum LDCP codes, as circuit-level decoding graphs can contain tens of thousands of nodes~\cite{bravyi_high-threshold_2024}.
Even with specialized hardware, inverting the matrix of a graph of this size cannot realistically be achieved within the decoherence time of a typical qubit~\cite{valls_syndrome-based_2021}. Whilst the BP+OSD decoder is a useful tool for simulations, it is not generally considered a practical method for real-time decoding.

In this work, we introduce \textit{localized statistics decoding} (LSD) as a parallel and efficient decoder for QLDPC codes, designed specifically to address the aforementioned limitations of BP+OSD, while retaining generality and good decoding performance. 
The key idea underpinning LSD is that in the sub-threshold regime, errors typically span disconnected areas of the decoding graph. 
Instead of inverting the entire decoding graph, LSD applies matrix inversion independently and concurrently for the individual \textit{sub-graphs} associated with these decoding regions. 
Similar to OSD, the performance of LSD can be improved using the soft information output of a pre-decoder such as BP. 
Our numerical decoding simulations of surface codes, bicycle bivariate codes, and hypergraph product codes show that our implementation of the BP+LSD decoder performs on par with BP+OSD in terms of decoding performance.

The efficiency of the LSD algorithm is made possible by a new linear algebra routine, which we call \textit{on-the-fly elimination}, that transforms the serial process of Gaussian elimination into a parallel one. Specifically, our method allows separate regions of the decoding graph to be reduced on separate processors. 
A distinct feature of on-the-fly elimination lies in a sub-routine that efficiently manages the extension and merging of decoding regions without necessitating the re-computation of row operations. 
The methods we introduce promise reduced runtime in the sub-threshold regime and open the possibility of using inversion-based decoders to decode real syndrome information from quantum computing experiments. We anticipate that on-the-fly elimination will also find broader utility in efficiently solving sparse linear systems across various settings, such as recommender systems~\cite{koren_matrix_2009} or compressed sensing~\cite{boche2015survey}.

\section{Results}

\subsection{The decoding problem} %
In this paper, we focus on the \emph{Calderbank-Shor-Steane} (CSS) subclass of QLDPC codes. 
These codes are defined by constant weight Pauli-$X$ and -$Z$ operators called \emph{checks} that generate the stabilizer group defining the code space.
In a gate-based model of computation, the checks are measured using a circuit containing auxiliary qubits and two-qubit Clifford gates that map the expectation value of each check onto the state of an auxiliary qubit.
The circuit that implements all check measurements is called the \emph{syndrome extraction circuit}.

For the decoding of QLDPC codes, the decoder is provided with a matrix $H\in\Ft^{|D|\times |F|}$ called the \textit{detector check matrix}. 
This matrix maps circuit fault locations $F$ to so-called \textit{detectors} $D$, defined as linear combinations of check measurement outcomes that are deterministic in the absence of errors. 
Specifically, each row of $H$ corresponds to a detector and each column to a fault, and $H_{df}=1$ if fault $f\in 
\{1,\dots,|F|\}$ flips detector $d\in \{1,\dots,|D|\}$. 
Such a check matrix can be constructed by tracking the propagation of errors through the syndrome extraction circuit using a stabilizer simulator~\cite{gidney_stim_2021,fang_symphase_2023,derks2024designing}.

We emphasize that, once the detector matrix $H$ is created, the minium-weight decoding problem can be mapped to the problem of decoding a classical linear code:
Given a \emph{syndrome} $\vec{s}\in \Ft^{|D|}$, the \emph{decoding problem} consists of finding a minimum-weight recovery $\hat{\mathbf{e}}$ such that $\vec{s}=H\cdot\hat{\vec{e}}$, where the vector $\hat{\mathbf{e}} \in \Ft^{|F|}$ indicates the locations in the circuit where faults have occurred.

The \emph{decoding graph} is a bipartite graph \mbox{$\mathcal{G}(H)=(V_D \cup V_F, E)$} with \emph{detector} nodes $V_D$, \emph{fault} nodes $V_F$ and edges $(d,f) \in E \iff H_{df}=1$.
$\mathcal{G}(H)$ is also known as the Tanner graph of the check matrix $H$.
Since the detector check matrix is analogous to a parity check matrix of a classical linear code, we use the terms detectors and checks synonymously.
Note that we implement a minimum-weight decoding strategy where the goal is to find the lowest-weight error compatible with the syndrome. This is distinct from maximum-likelihood decoding, where the goal is to determine the highest probability logical coset.

\subsection{Localized statistics decoding}
This section provides an example-guided outline of the localized statistics decoder. A more formal treatment, including pseudo-code, can be found in~\Cref{sec:lsd_algorithm}.
\begin{figure}[!b]
    \centering
    \includegraphics{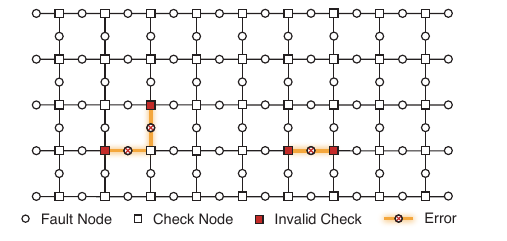}
    \caption{Illustration of the factorization of the decoding problem on a $5 \times 10$ surface code patch. Below the threshold, errors are typically sparsely distributed on the decoding graph and form small clusters with disjoint support.}
    \label{fig:matrix_factorization}
\end{figure}
\paragraph{Notation.} For an index set $I = \{i_1, \dots, i_n\}$ and a matrix $M = (m_1, \dots, m_\ell)$ with columns $m_j$, we write $M_{[I]} = (m_{i_1}, \dots, m_{i_n})$ as the matrix containing only the columns indexed by $I$.
Equivalently, for a vector $\vec{v}$, $\vec{v}_{[I]}$ is the vector containing only coordinates indexed by $I$. 

\paragraph{Inversion decoding.} The 
\emph{localized statistics decoding} (LSD) algorithm belongs to the class of reliability-guided inversion decoders, which also contains \emph{ordered statistics decoding} (OSD)~\cite{fossorier_reliability-based_1998, panteleev_degenerate_2021, roffe_decoding_2020}. 
OSD can solve the decoding problem by computing $\vec{\hat{e}}_{[I]}=H_{[I]}^{-1}\cdot \vec{s}$.
Here, $H_{[I]}$ is an invertible matrix formed by selecting a linearly independent subset of the columns of the check matrix $H$ indexed by the set of column indices $I$. 
The algorithm is reliability-guided in that it uses prior knowledge of the error distribution to strategically select $I$ so that the solution $\vec{\hat{e}}_{[I]}$ spans faults that have the highest error probability. 
The reliabilities can be derived, for example, from the device's physical error model~\cite{pattison_improved_2021, raveendran2022soft, berent_analog_2024, bourassa_blueprint_2021, tzitrin_fault-tolerant_2021, vuillot_quantum_2019} or the soft information output of a \textit{pre-decoder} such as BP~\cite{higgott_improved_2023}.

\paragraph{Factorizing the decoding problem.} In general, solving the system $\vec{\hat{e}}_{[I]}=H_{[I]}^{-1}\cdot \vec{s}$ involves applying Gaussian elimination to compute the inverse $H_{[I]}^{-1}$, which has cubic worst-case time complexity, $O(n^3)$, in the size $n$ of the check matrix $H$.
The essential idea behind the LSD decoder is that, for low physical error rates, the decoding problem for QLDPC amounts to solving a sparse system of linear equations.
In this setting, the inversion decoding problem can be factorized into a set of independent linear sub-systems that can be solved concurrently.

\Cref{fig:matrix_factorization} shows an example of error factorization in the Tanner graph of a $5\times10$ surface code. 
The support of a fault vector $\vec{e}$ is illustrated by the circular nodes marked with an $X$ and the corresponding syndrome is depicted by the square nodes filled in red. 
In this example, it is clear that $\vec{e}$ can be split into two connected components, $\vec{e}_{[C_1]}$ and $\vec{e}_{[C_2]}$, that occupy separate regions of the decoding graph. 
We refer to each of the connected components induced by an error on the decoding graph as \emph{clusters}. 
With a slight misuse of notation, we refer to clusters $C_i$ and their associated incidence matrices $H_{[C_i]}$ interchangeably and use $|C_i|$ to denote the number of fault nodes (columns) in the cluster (its incidence matrix, respectively). 
This identification is natural as clusters are uniquely identified by their fault nodes, or equivalently, by column indices of $H$: for a set of fault nodes $C \subseteq V_F$, we consider all of the detector nodes in $V_D$ adjacent to at least one node in $C$ to be part of the cluster. 

For the example in~\figref{fig:matrix_factorization}, the two induced clusters $H_{[C_1]}, H_{[C_2]}$ are entirely independent of one another. 
As such, it is possible to find a decoding solution by inverting each submatrix separately,
\begin{equation}
    \vec{\hat{e}}_{[C_1\cup C_2\cup C_{\bot}]}=\left(H_{[C_1]}^{-1} \vec{s}_{[C_1]}, H_{[C_2]}^{-1} \vec{s}_{[C_2]},  0 \right),
\end{equation}
where $\vec{s}_{[C_{i}]}$ is the subset of syndrome bits in the cluster $H_{[C_i]}$, which we refer to as the \textit{cluster syndrome}. 
The set $C_{\bot}$ is the column index set of fault nodes that are not in any cluster.

In general, linear systems can be factorized into $\nu$ many decoupled clusters, yielding
\begin{equation}\label{eq:lsd_factorization}
    \vec{\hat{e}}_{[C_1\cup \dots\cup C_\nu\cup C_\bot]}
    =\left(H_{[C_1]}^{-1} \vec{s}_{[C_1]}, \dots,  H_{[C_\nu]}^{-1} \vec{s}_{[C_\nu]},  0 \right).
\end{equation}
The number of clusters, $\nu$, will depend upon $H$, the physical error rate, and the Hamming weight of $\mathbf{s}$. 
If a factorization can be found, matrix inversion is efficient: first, the $\nu$ clusters can be solved in parallel; second, the parallel worst-case time complexity of the algorithm depends on the maximum size of a cluster \mbox{$\kappa=\text{max}_i\left(|C_i|\right)$}, where $|C_i|$ is the number of fault nodes in $C_i$. The worst-case scaling $O(\kappa^3)$ contrasts with the $O(n^3)$ OSD post-processing scaling, where $n = |V_F|$ is the size of the matrix $H$. 
To enable parallel execution, we have devised a routine that we call \emph{on-the-fly elimination} to efficiently merge clusters and compute a matrix factorization, as detailed in~\Cref{sec:fly_solve}.
\begin{figure*}[!]
    \centering
    \includegraphics{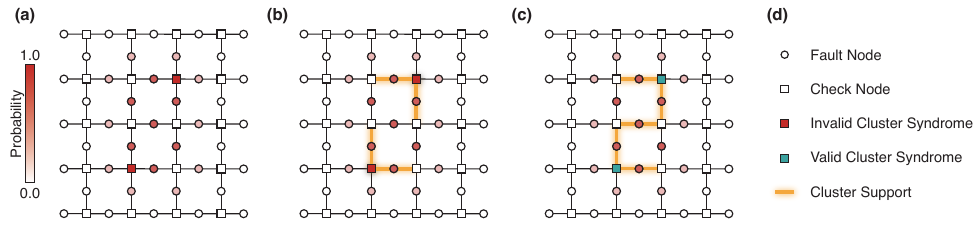}
    \caption{Reliability-based weighted cluster growth example for the surface code. 
    \textbf{(a)} The syndrome of an error is indicated as red square vertices.
    The fault nodes are colored to visualize their error probabilities obtained from belief propagation pre-processing.
    \textbf{(b)} Clusters after the first two growth steps. 
    In the guided cluster growth strategy, fault nodes are added individually to the local clusters. 
    The order of adding the first two fault nodes to each cluster is random since both have the same probability due to the presence of degenerate errors.
    \textbf{(c)} After an additional growth step, the two clusters are merged and the combined cluster is valid. 
    \textbf{(d)} Legend for the used symbols.}
    \label{fig:weighted-growth}
\end{figure*}
\paragraph{Weighted cluster growth and the LSD validity condition.}
For a given syndrome, the LSD algorithm is designed to find a factorization of the decoding graph that is as close to the optimal factorization as possible. Here, we define a factorization as \textit{optimal} if its clusters correspond exactly to the connected components induced by the error.

The LSD decoder uses a weighted, reliability-based growth strategy to factorize the decoding graph. 
The algorithm begins by creating a cluster $H_{[C_i]}$ for each flipped detector node, i.e., a separate cluster is generated for every nonzero bit in the syndrome vector $\vec{s}$. 
At every growth step, each cluster $H_{[C_i]}$ is grown by one column by adding the fault node from its neighborhood with the highest probability of being in error according to the input reliability information.
This weighted growth strategy is crucial for controlling the cluster size: limiting growth to a single fault node per time step increases the likelihood that an efficient factorization is found, especially for QLDPC codes with high degrees of expansion in their decoding graphs. 

If two or more clusters collide -- that is, if a check node would be contained in multiple clusters after a growth step  -- the LSD algorithm merges them and forms a combined cluster. 
We use the notation $H_{[C_1 \cup C_2]}$ and $\vec{s}_{[C_1\cup C_2]}$ to indicate the decoding matrix and the syndrome of the combined cluster.

For each cluster $C_i$, the LSD algorithm iterates cluster growth until it has enough linearly independent columns to find a local solution, i.e., until \mbox{$\mathbf{s}_{[C_i]}\in \text{image}(H_{[C_i]})$}. 
We call such a cluster \emph{valid}. 
Once all clusters are valid, the LSD algorithm computes all local solutions, $\mathbf{e}_{[C_i]}=H_{[C_i]}^{-1}\cdot \mathbf{s}_{[C_i]}$, and combines them into a global one.

The process of weighted-cluster growth is conceptually similar to ``belief hypergraph union-find''~\cite{cain2024correlated} and is illustrated in \figref{fig:weighted-growth} for the surface code. 
Here, two clusters are created.
These are grown according to the reliability ordering of the neighboring fault nodes. 
In~\sfigref{fig:weighted-growth}{c}, the two clusters merge, yielding a combined valid cluster. 
The combined cluster is not optimal as its associated decoding matrix has $5$ columns, whereas the local solution has Hamming weight $3$, indicating that the optimal cluster would have $3$ columns. 
Nonetheless, computing a solution using the cluster matrix with $5$  columns is still preferable to computing a solution using the full $41$-column decoding matrix -- this highlights the possible computational gain of LSD.

\paragraph{\emph{On-the-fly} elimination and parallel implementation.}
To avoid the overhead incurred by checking the validity condition after each growth step -- a bottleneck for other clustering decoders for QLDPC codes~\cite{delfosse_toward_2021} -- we have developed an efficient algorithm that we call \emph{on-the-fly elimination}.
Our algorithm maintains a dedicated data structure that allows for efficient computation of a matrix factorization of each cluster when additional columns are added to the cluster, even if clusters merge -- see \secref{sec:fly_solve} for details.
Importantly, at each growth step, due to our on-the-fly technique, we only need to eliminate a single additional column vector without having to re-eliminate columns from previous growth steps.

Crucially, on-the-fly elimination can be applied in parallel to each cluster $H_{[C_i]}$. 
Using the on-the-fly data structure that enables clusters to be efficiently extended without having to recompute their new factorization from scratch, we propose a fully \emph{parallel implementation of LSD} in \appref{sec:plsd}.
There, we analyze parallel LSD time complexity and show that the overhead for each parallel resource is low and predominantly depends on the cluster sizes.

\paragraph{Factorization in decoding graphs.} 
A key feature of LSD is to divide the decoding problem into smaller, local sub-problems that correspond to error clusters on the decoding graph.
To provide more insight, we investigate cluster formation under a specific noise model and compare these clusters obtained directly from the error to the clusters identified by LSD.

\begin{figure}[tb]
    \centering
    \includegraphics[width=1.\columnwidth]{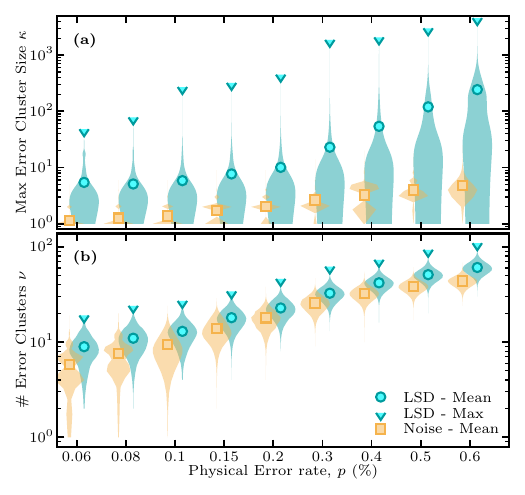}
    \caption{Cluster size statistics of the $\llbracket 144, 12, 12 \rrbracket$ bivariate bicycle code of Ref.~\cite{bravyi_high-threshold_2024} under circuit-level noise with strength $p$. Markers show the mean of the distribution while shapes are violin plots of the distribution obtained from $10^5$ samples. Yellow distributions show statistics for the optimal factorization while the blue distributions show statistics for the factorization returned by BP+LSD.
    We show in panel \textbf{(a)} the distribution of the maximum cluster size $\kappa$ and in \textbf{(b)} the distribution of the cluster count, $\nu$, for each decoding sample.
    Markers and distributions are slightly offset from the actual error rate to increase readability.}
    \label{fig:cluster_size_stats}
\end{figure}

As a timely example, we focus specifically on the cluster size statistics of the circuit-noise decoding graph of the $\llbracket 144,12,12 \rrbracket$ bivariate bicycle code~\cite{kovalev_quantum_2013} that was recently investigated in Ref.~\cite{bravyi_high-threshold_2024}.
\sfigref{fig:cluster_size_stats}{a} shows the distribution of the maximum sizes of clusters identified by BP+LSD over $10^5$ decoding samples, see~\cref{ssec:noise_model} for details.
The figure illustrates that for low enough noise rates, the largest clusters found by LSD are small and close to the optimal sizes of clusters induced by the original error, even if only a relatively small number (30) of BP iterations is used to compute the soft information input to LSD.
It is worth emphasizing that large clusters are typically formed by merging two, or more, smaller clusters identified and processed at previous iterations of the algorithm. Owing to our on-the-fly technique that processes the linear system corresponding to each of these clusters (cf.~\Cref{sec:fly_solve}), the maximum cluster size only represents a loose upper bound on the complexity of the LSD algorithm.
 
\sfigref{fig:cluster_size_stats}{b} shows the distributions of the cluster count per shot, $\nu$ -- that is, per shot, where the LSD data is post-selected on shots where BP does not converge -- against the physical error rate $p$.
The number of clusters $\nu$ corresponds to the number of terms in the factorization of the decoding problem and thus indicates the degree to which the decoding can be parallelized, as disjoint factors can be solved concurrently. At practically relevant error rates below the (pseudo) threshold, e.g.,  $p \leq 0.1 \%$, we observe on average 10 independent clusters. This implies that the LSD algorithm benefits from parallel resources throughout its execution.

We explore bounds on the sizes of clusters induced by errors on QLDPC code graphs in \appref{app:qldpc_clusters}.
Our findings suggest that detector matrices generally exhibit a strong suitability for factorization, a feature that the LSD algorithm is designed to capitalize on.
\paragraph{Higher-order reprocessing.}

Higher-order reprocessing in OSD is a systematic approach designed to increase the decoder's accuracy.
The zero-order solution $\vec{\hat{e}}_{[I]}=H_{[I]}^{-1}\cdot \vec{s}$ of the decoder cannot be made lower weight if the set of column indices $I$ specifying the invertible submatrix $H_{[I]}$ matches the $\vert I \vert$ most likely fault locations identified from the soft information vector $\vec{\lambda}$.
However, if there are linear dependencies within the columns formed by the $\vert I \vert$ most likely fault locations, the solution $\hat{\vec{e}}$ may not be optimal. 
In those cases, some fault locations in $\overline{I}$ (the complement of $I$) might have higher error probabilities.
To find the optimal solution, one can systematically search all valid fault configurations in $\overline{I}$ that potentially provide a more likely estimate $\hat{\vec{e}}'$.
This search space, however, is exponentially large in $\vert \overline{I} \vert$.
Thus, in practice, only configurations with a Hamming weight up to $w$ are considered, known as \emph{order$-w$} reprocessing.
See Refs.~\cite{fossorier_soft-decision_1995, fossorier_reliability-based_1998, panteleev_degenerate_2021, roffe_decoding_2020} for a more technical discussion.

In BP+OSD-$w$ applied to $H$, order-$w$ reprocessing is frequently the computational bottleneck because of the extensive search space and the necessary matrix-vector multiplications involving $H_{[I]}$ and $H_{[\overline{I}]}$ to validate fault configurations. 
Inspired by higher-order OSD, we propose a higher-order reprocessing method for LSD, which we refer to as LSD$-\mu$.
We find that when higher-order reprocessing is applied to LSD, it is sufficient to process clusters locally.
This offers three key advantages: parallel reprocessing, a reduced higher-order search space, and smaller matrix-vector multiplications.
Furthermore, our numerical simulations indicate that decoding improvements of local BP+LSD$-\mu$ are on par with those of global BP+OSD$-w$.
For more details on higher-order reprocessing with LSD and additional numerical results, see~\appref{app:higher_order_lsd}.

\begin{figure*}[tbh]
	\centering
	\includegraphics{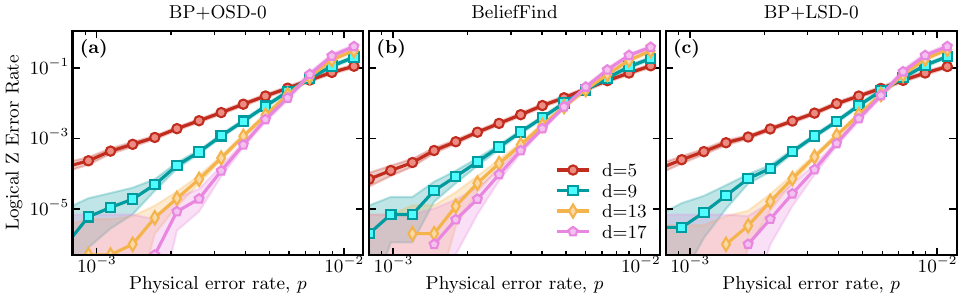}
	\caption{Comparison of various decoders guided by belief propagation for decoding rotated surface codes of distance $d$ subject to circuit-level depolarizing noise parameterized by a single parameter, called the \emph{physical error rate} $p$, see~\Cref{ssec:noise_model} for details.
		We use \texttt{Stim} to perform a \texttt{surface\_code:rotated\_memory\_z} experiment for $d$ syndrome extraction cycles with single and two-qubit error probabilities $p$.
		\textbf{(a)} The performance of BP+OSD-0 that matches the performance of the proposed decoder.
		\textbf{(b)} The performance of a BeliefFind decoder that shares a cluster growth strategy with the proposed decoder.
		\textbf{(c)} Performance of the proposed BP+LSD decoder.
		The shading indicates hypotheses whose likelihoods are within a factor of 1000 of the maximum likelihood estimate, similar to a confidence interval.
	}
	\label{fig:sc_ckt_stim}
\end{figure*}

\subsection{Numerical results} \label{ssec:numerical_results}
For the numerical simulations in this work, we implement serial LSD, where the reliability information is provided by a BP pre-decoder. The BP decoder is run in the first instance, and if no solution is found, LSD is invoked as a post-processor. Our serial implementation of this BP+LSD decoder is written in C\texttt{++} with a python interface and is available open-source as part of the \texttt{LDPCv2} package~\cite{ldpc_main}.

Our main numerical finding is that BP+LSD can decode QLDPC codes with performance on par with BP+OSD.
We include the results of extensive simulations in which BP+LSD is used to decode a circuit-level depolarizing noise model for surface codes, hypergraph product 
 (HGP) codes~\cite{tillich_quantum_2014}, and bivariate bicycle codes~\cite{bravyi_high-threshold_2024, kovalev_quantum_2013}.

In BP+OSD decoding, it is common to run many BP iterations to maximize the chance of convergence and reduce the reliance on OSD post-processing.
A strength of the BP+LSD decoder is that LSD is less costly than OSD and, therefore, applying the LSD routine after running BP introduces comparatively small overall computational overhead. 
As a result, the number of BP iterations in BP+LSD can be considerably reduced since LSD requires only a few BP iterations to obtain meaningful soft information values.
This is in stark contrast to BP+OSD, where it is often more efficient to run many BP iterations rather than deferring to costly OSD. 
In this work, we use a fixed number of $30$ BP iterations for all decoding simulations with BP+LSD. 
For context, this is a significant reduction compared to the decoding simulations of Ref.~\cite{bravyi_high-threshold_2024} where BP+OSD was run with $10^4$ BP iterations.

\paragraph{Surface codes.}
We compare the threshold of BP+LSD with various state-of-the-art decoders that are similarly guided by the soft information output of a BP decoder. 
In particular, we compare the proposed BP+LSD algorithm with BP+OSD (order 0)~\cite{roffe_decoding_2020}, as well as our implementation of a BP plus union-find (BP+UF) decoder~\cite{higgott_improved_2023-1} that is tailored to matchable codes.
The results are shown in~\figref{fig:sc_ckt_stim}.
The main result is that both BP+OSD and BP+LSD achieve a similar threshold close to a physical error rate of $p \approx 0.7\%$, and similar logical error rates, see panels (a) and (c), respectively. 
In particular, in the relevant sub-threshold regime, where BP+LSD can be run in parallel, its logical decoding performance matches BP+OSD.
Note that this is the desired outcome and demonstrates that our algorithm achieves (close to) identical performance with BP+OSD while maintaining locality.
Our implementation of the BP+UF decoder of Ref.~\cite{higgott_improved_2023}, see panel (b), performs slightly worse, achieving a threshold closer to $p \approx 0.6\%$ and higher logical error rates, potentially due to a non-optimized implementation.

\begin{figure}[!b]
	\centering
	\includegraphics{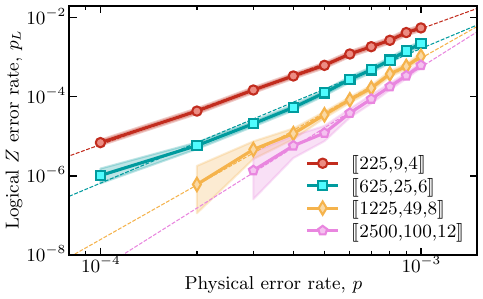}
	\caption{Below threshold logical error rate $p_L$ of a family of $\llbracket 25s^2, s^2 \rrbracket$ constant-rate hypergraph product codes decoded with of the BP+LSD decoder. We simulate $N_c = 12$ rounds of syndrome extraction cycles under circuit-level noise with physical error rate $p$ and apply a $(3, 1)-$overlapping window technique to enable fast and accurate single-shot decoding, see~\Cref{ssec:noise_model} for details. The shading indicates hypotheses whose likelihoods are within a factor of 1000 of the maximum likelihood estimate, similar to a confidence interval. Dashed lines are an exponential fit
    with a linear exponent to the numerically observed error
    rates.}
	\label{fig:harvard_codes_owd}
\end{figure}

\paragraph{Random (3,4)-regular hypergraph product codes.}
\figref{fig:harvard_codes_owd} shows the results of decoding simulations for the family of hypergraph product codes~\cite{tillich_quantum_2014} that were recently studied in Ref.~\cite{xu_constant-overhead_2024}. The plot shows the logical error rate per syndrome cycle $p_L = 1 - (1 - P_L(N_c))^{1/N_c}$, where $N_c$ is the number of syndrome cycles and $P_L(N_c)$ the logical error rate after $N_c$ rounds.
Under the assumption of an identical, independent circuit-level noise model, BP+LSD significantly outperforms the BP plus small set-flip (BP+SSF) decoder investigated in Ref.~\cite{tremblay_constant-overhead_2021}.
For example, for the $\llbracket 625, 25 \rrbracket$ code instance at $p\approx0.1\%$, BP+LSD improves logical error suppression by almost two orders of magnitude compared to BP+SSF.
\paragraph{Bivariate bicycle codes.}
\begin{figure}[t]
	\centering
	\includegraphics{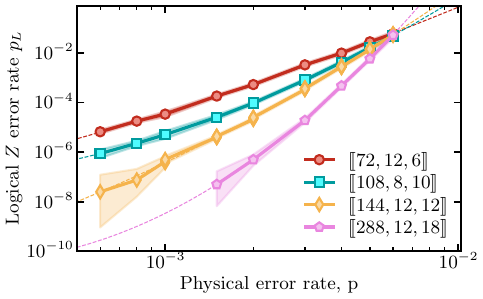}
	\caption{Logical error rate per syndrome cycle $p_L$ for various bivariate bicycle codes under a circuit-level noise model. For each code, $d$ rounds of syndrome extraction are simulated and the full syndrome history is decoded using BP+LSD.
    The shading highlights the region of estimated probabilities where the likelihood ratio is within a factor of 1000, similar to a confidence interval. Dashed lines are an exponential fit with a quadratic exponent to the numerically observed error rates.}
	\label{fig:ibm_codes_full}
\end{figure}
Here, we present decoding simulation results of the \emph{bivariate bicycle} (BB) codes.
These codes are part of the family of hyperbicycle codes originally introduced in Ref.~\cite{kovalev_quantum_2013}, and more recently investigated at the circuit level in Ref.~\cite{bravyi_high-threshold_2024}.
In~\figref{fig:ibm_codes_full}, we show the logical error $Z$ rate per syndrome cycle, $p_{L_{Z}}$.
We find that with BP+LSD we obtain comparable decoding performance to the results presented in Ref.~\cite{bravyi_high-threshold_2024} where simulations were run using BP+OSD-CS-7 (where BP+OSD-CS-7 refers to the ``combination sweep'' strategy for BP+OSD higher-order processing with order $w=7$, see Ref.~\cite{roffe_decoding_2020} for more details).

\paragraph{Runtime statistics.}
To estimate the time overhead of the proposed decoder in numerical simulation scenarios and to compare it with a state-of-the-art implementation of BP+OSD, we present preliminary timing results for our prototypical open-source implementation of LSD in~\appref{sec:runtime-estimation}. We note that for a more complete assessment of performance, it will be necessary to benchmark a fully parallel implementation of the algorithm, designed for specialized hardware such as GPUs or FPGAs. We leave this as a topic for future work.

\section{Discussion}
When considering large QLDPC codes, current state-of-the-art decoders such as BP+OSD hit fundamental limitations due to the size of the resulting decoding graphs. 
This limitation constitutes a severe bottleneck in the realization of protocols based on QLDPC codes.
In this work, we address this challenge through the introduction of the LSD decoder as a parallel algorithm whose runtime depends predominantly on the physical error rate of the system. 
Our algorithm uses a reliability-based growth procedure to construct clusters on the decoding graph in a parallel fashion. 
Using a novel routine that computes the PLU decomposition~\cite{strang2022introduction} of the clusters' sub-matrices on-the-fly, we can merge clusters efficiently and compute local decoding solutions in a parallel fashion. 
Our main numerical findings are that the proposed decoder performs on par with current state-of-the-art decoding methods in terms of logical decoding performance.

A practical implementation of the algorithm has to be runtime efficient enough to overcome the so-called backlog problem~\cite{terhal_quantum_2015}, where syndrome data accumulates since the decoder is not fast enough. 
While we have implemented an overlapping window decoding technique for our algorithm, it might be interesting to further investigate the performance of LSD under parallel window decoding~\cite{skoric_parallel_2023}, where the overlapping decoding window is subdivided to allow for further parallelization of syndrome data decoding.

To decode syndrome data from quantum computing experiments in real-time, it will be necessary to use specialized hardware such as 
\emph{field programmable gate arrays} (FPGAs) or \emph{application-specific integrated circuits} (ASICs), as recently demonstrated for variants of the union-find surface code decoder~\cite{barber2023realtime,liyanage2024fpgabased,das2022afs} or possibly cellular automaton based approaches \cite{CADecoders}.
A promising avenue for future research is to explore the implementation of an LSD decoder on such hardware to assess its performance with real-time syndrome measurements.

Concerning alternative noise models, erasure-biased systems have recently been widely investigated~\cite{kubica2023erasure, levine2023demonstrating,gu2023fault,wu2022erasure}. 
We conjecture that LSD can readily be generalized to erasure decoding, either by adapting the cluster initialization or by considering a re-weighting procedure of the input reliabilities. 
We leave a numerical analysis as a topic for future work.

Finally, it would be interesting to investigate the use of maximum-likelihood decoding at the cluster level as recently explored in Ref.~\cite{wolanski_ambiguity_2024} as part of the BP plus ambiguity clustering (BP+AC) decoder. Specifically, such a method could improve the efficiency of the LSD$-\mu$ higher-order reprocessing routines we explored. Similarly, the BP+AC decoder could benefit from the results of this paper: our parallel LSD cluster growth strategy, combined with on-the-fly elimination, provides an efficient strategy for finding the BP+AC block structure using parallel hardware.

\section{Methods}
\subsection{LSD algorithm} \label{sec:lsd_algorithm}
In this section, we provide a detailed description of the LSD algorithm and its underlying data structure designed for efficient cluster growth, merging, validation, and ultimately local inversion decoding. 
We start with some foundational definitions.

\begin{definition}[Clusters]\label{def:cluster}
    Let $\mathcal{G}(H) = (V_D \cup V_F, E)$ be the decoding graph of a QLDPC code with detector nodes $V_D$ and fault nodes $V_F$.
    There exists an edge $(d,f) \in E \iff H_{df}=1$.
    A \emph{cluster} \mbox{$C = (V^C_D \cup V^C_F,E^C) \subseteq \mathcal{G}(H)$} is a connected component of the decoding graph.

\end{definition}

\begin{definition}
[Cluster sub-matrix]\label{def:cluster_sub-matrix}    
    Given a set of column indices $C$ of a cluster, the sub-matrix $H_{[C]}$ of the check matrix $H$ is called the \emph{cluster sub-matrix}.
        The \emph{local syndrome} $\vec{s}_{[C]}$ of a cluster is the support vector of detector nodes in the cluster.
        A cluster is \emph{valid} if $\vec{s}_{[C]}\in \text{\textsc{image}}(H_{[C]})$. Note that a cluster is uniquely identified by the columns of its sub-matrix $H_{[C]}$, hence we use $H_{[C]}$ to denote both the cluster and its sub-matrix.
\end{definition}

\begin{definition}
    [Cluster-boundary and candidate fault nodes]\label{def:cluster_boundary}
    The set of \emph{boundary detector} nodes $\beta(C) \subseteq V_D^C$ of a cluster $C$ 
    is the set
    \begin{equation}
        \beta(C) = \{d \in V_D^C \mid \Gamma(d) \not \subseteq V_F^C\} 
    \end{equation}
    of all detector nodes in $C$ that are connected to at least one fault node not in $C$,
    where $\Gamma(v)$ is the neighborhood of the vertex v, i.e., $\Gamma(v) = \{ u \in \mathcal{G}(H) \mid (v, u) \in E\}$. 
    We define \emph{candidate fault nodes} $\Lambda(C) \subseteq V_F \setminus V_F^C$ as the set of fault nodes not in $C$ and connected to at least one boundary detector node in $\beta(C)$
    \begin{equation}
        \Lambda(C) = \Gamma\left(\beta(C)\right) \cap \left(V_F \setminus V_F^C
        \right).
    \end{equation}
\label{def:boundary_cluster}
\end{definition}

\begin{definition}
    [Cluster collisions] \label{def:cluster_collisions}
    Two or more clusters $\{ C_i \}$ \emph{collide} due to a set of fault nodes $\Delta_F$ if 
    \begin{align}
    \Delta_F \subseteq \bigcup_{i} \Lambda(C_i) \text{\quad and\quad } 
    \bigcap_i \beta(C_i) \cap \Gamma(\Delta_F) \neq \emptyset.
    \end{align}
\end{definition}

The LSD algorithm takes as input the matrix \mbox{$H \in \Ft^{m\times n}$}, where $m=|V_D|, \ n=|V_F|$, a syndrome $\vec{s} \in \Ft^{m}$, and a reliability vector that contains the soft information $\vec{\lambda} \in \R^n$. 
In the following, we will assume that $\vec{\lambda}$ takes the form of \emph{log-likelihood-ratios} (LLRs) such that the lower the LLR, the higher the probability that the corresponding fault belongs to the error. 
For instance, this is the form of soft information that is returned by the BP decoder.

A sequential version of the algorithm is outlined below and detailed in the pseudo-code in~\Cref{alg:bp_lsd_basic}. 
A parallel version of the LSD algorithm is presented in~\appref{sec:plsd}.
\begin{enumerate}
    \item A cluster is created for each flipped detector node $d_i$ where $\vec{s_i}=1$. 
    This cluster is represented by its corresponding sub-matrix $H_{[C_i]}$. 
    Initially, each cluster is added to a list of invalid clusters. 
    \item Every cluster is grown by a single node $v_j$ drawn from the list of candidate nodes $\Lambda(C_i)$. For the first growth step after cluster initialization, we define $\Lambda(C_i)= \Gamma(\{s_i\}) $ --  see \cref{def:boundary_cluster}.
    The chosen growth node $v_j\in\Lambda(C_i)$ in each step is the fault node with the highest probability of being in error. 
    That is, $v_j$ has the lowest value among the LLRs for the candidate fault nodes $\lambda_j < \lambda_{j+1} < \dots < \lambda_{\ell}, \ \ell = |\Lambda(C)|$. 
    Hence, the growth step involves adding one new column to the cluster matrix $H_{[C_i]}$.
    \item During growth, the algorithm detects collisions between clusters due to the selected fault nodes. 
    Clusters that collide are merged.
    \item The Gaussian elimination row operations performed on previous columns are performed on the new column together with the row operations needed to eliminate the newly added columns of $H_{[C_i]}$. 
    In addition, every row operation applied to $H_{[C_i]}$ is also applied to the local syndrome $\vec{s}_{[C_i]}$. 
    This allows the algorithm to efficiently track when the cluster becomes valid. 
    Explicitly, the cluster is valid when the syndrome becomes linearly dependent on the cluster decoding matrix i.e., when $\vec{s}_{[C_i]}\in \text{\textsc{image}}(H_{[C_i]})$. 
    In addition to cluster validation, the Gaussian elimination at each step enables an on-the-fly computation of the PLU factorization of the local cluster. 
    We refer the reader to \cref{sec:fly_solve} for an outline of our on-the-fly elimination method. 
    \item The valid clusters are removed from the invalid cluster list, and the algorithm continues iteratively until the invalid cluster list is empty. 
    \item Once all clusters are valid, the local solutions $\vec{\hat{e}}_{[C_i]}$ such that $H_{[C_i]}\cdot \vec{\hat{e}}_{[C_i]} = \vec{s}_{[C_i]}$ can be computed via the PLU decomposition of each cluster matrix $H_{[C_i]}$ that has been computed on-the-fly during cluster growth. 
    The output of the LSD algorithm is the union of all the local decoding vectors $\vec{\hat{e}}_{[C_i]}$.
\end{enumerate}
\newcommand\mycommfont[1]{\footnotesize\ttfamily\textcolor{black}{#1}}
\SetCommentSty{mycommfont}
\begin{algorithm}[tbh]
\SetAlgoLined
\DontPrintSemicolon
 \caption{Localized statistics decoding (LSD) -- serial algorithm \label{alg:bp_lsd_basic}} %
    $H$: decoding matrix\;
    $\vec{s}$: syndrome vector\;
    $\vec{\lambda}$: fault node soft information vector\;
    $\invalid \coloneqq [\,]$: list of invalid clusters $\mathrm{cl_i}$\;
    $\valid \coloneqq [\,]$: list of valid clusters $\mathrm{cl_\ell}$\;
    \For{$s_i \in \mathbf{s}$}{
        $\mathrm{cl_i}$ = \textsc{create\_cluster}($s_i$)\;
       $\invalid$.add($\mathrm{cl_i}$)\;
    }
    \While{$\invalid\neq [\,]$} { \label{alg:line:invalidClusters}
        \For{$\mathrm{cl} \in \invalid$}{
            cl.\textsc{grow\_cluster}($\vec{\lambda}$)\;
        }

        \For{$\mathrm{cl_i}, \mathrm{cl_{\Cjindex}} \in \invalid\cup \valid$}{ \tcp*[l]{check if any clusters $\mathrm{cl_i}, \mathrm{cl_\ell}$ collide}
            merged = \textsc{check\_collision}$(\mathrm{cl_i}, \mathrm{cl_{\Cjindex}})$\;
            \If{{\text{merged}}}{
                $\mathrm{cl_{i\cup\Cjindex}}=$ \textsc{merge\_clusters}$(\mathrm{cl_i}, \mathrm{cl_{\Cjindex}})$\;
                $\invalid$.remove($\{\mathrm{cl_i}, \mathrm{cl_{\Cjindex}}\}$)\;
                $\valid$.remove($\{\mathrm{cl_i}, \mathrm{cl_{\Cjindex}}\}$)\;
                $\invalid$.add($\mathrm{cl_{i\cup \Cjindex}}$)\;
            }
        }

        \For{$\mathrm{cl} \in \invalid$}{
            cl.\textsc{plu\_decompose()}\;
            valid = cl.\textsc{check\_validity}($\vec{s}_{[\mathrm{cl}]}$)\;
            \If{{ $\text{valid}$}}{
                $\valid$.add(cl)\;
                $\invalid$.remove(cl)\;
            } 
        }
   
    }
local\_decodings = [\,]\;
    \For{{$\text{cl} \in \valid$ }}{
        local\_decodings.append(cl.\textsc{plu\_solve($\vec{s}_{[\mathrm{cl}]})$)}\;
    }
    
    return \textsc{global\_decoding}(local\_decodings)\;
\end{algorithm}

\subsection{On-the-fly elimination}\label{sec:fly_solve}
A common method for solving linear systems of equations is to use a matrix factorization technique. 
A foundational theorem in linear algebra states that every invertible matrix $A$ factorizes as $A=PLU$, that is, there exist matrices $P,L,U$ such that 
\begin{equation}
	A = PLU,
\label{eq:plu}
\end{equation}
where $P$ is a permutation matrix, $U$ is upper triangular, and $L$ is lower triangular with $1$ entries on the diagonal.
Once in PLU form, a solution $x$ for the system $A\cdot x=y$ can be efficiently computed using the forward and back substitution procedure~\cite{strang2022introduction}. 
The computational bottleneck of this method to solve linear systems stems from the Gaussian elimination procedure required to transform $A$ into PLU form. 

Here, we present a novel algorithm called \textit{on-the-fly elimination} to efficiently compute the PLU factorization over $\Ft{}$.
Note that the algorithm can in principle be generalized to matrices over any field. However, in the context of coding theory, $\Ft{}$ is most relevant and we restrict the discussion to this case.

The main idea of the on-the-fly elimination is that row operations can be applied in a column-by-column fashion. 
If the operations that have been applied to each column of the matrix are stored, they can be applied to a newly added column such that only this column needs to be eliminated as all other columns are already in reduced form.
This highlights the nice interplay between cluster growth (i.e., appending columns) and the on-the-fly elimination for PLU factorization of the cluster matrix.

To grow and merge clusters, multiple smaller steps are necessary.
As detailed above, these steps include identifying fault nodes/column indices of the decoding matrix $H$ by which the invalid clusters will grow and determining whether an added fault node will lead to two or more clusters merging into a single one -- see~\Cref{def:cluster_collisions}.
For simplicity, we first describe the case of sequential cluster growth.
Our on-the-fly procedure can analogously be applied in a parallel implementation, see~\appref{sec:plsd}.

Let $C_i$ be an active cluster, that is, $\left(H_{[C_i]}, \vec{s}_{[C_i]}\right)$ does not define a solvable decoding problem as \mbox{$s_{[C_i]}\notin \textsc{image}(H_{[C_i]})$}. 
To grow cluster $C_i$, we consider candidate fault nodes $v_j \in \Lambda(C_i)$ -- fault nodes not already in $C_i$ but connected to check nodes on its boundary $\beta(C)$, see \cref{def:boundary_cluster}.
The candidate fault node with the highest probability of being in error according to the soft information vector $\vec{\lambda} \in \mathbb{R}^n$ is selected.
Once $v_j$ has been chosen, we check whether its neighboring detector nodes are boundary nodes of any other (valid or invalid) clusters i.e., we check for collisions, cf~\Cref{def:cluster_collisions}.
If this is not the case, we proceed as follows.
We now assume that the active cluster $C_i$ described by sub-matrix $H_{[C_i]}$ that has a PLU factorization of the form
\begin{align}
	H_{[C_i]} = P_{i} L_{i} U_{i},
\end{align}
where $P_i, L_i, U_i$ are as in~\cref{eq:plu}.
Adding a fault node $v_j$ to the cluster is equivalent to adding a (sparse) column vector $\vec{b}$ to $H_{[C_i]}$, i.e., 
\begin{align}
	H_{[C_i\cup \{v_j\}]} =\begin{pmatrix} \begin{array}{c|c}
		H_{[C_i]} &  \multirow{2}{*}{$ \vec{b}$} \\
		0 &
	\end{array} \end{pmatrix}.
 \label{eq:one_added_column}
\end{align}
A key insight is that the PLU factorization of the extended matrix $H_{[C_i \cup \{v_j\}]}$ can be computed through row operation on column $\vec{b}$ alone: it is not necessary to factorize the full matrix $H_{[C_i \cup \{v_j\}]}$ from scratch. 
By applying the PLU factorization of $H_{[C_i]}$ block-wise to the extended matrix $H_{[C_i \cup \{v_j\}]}$, we obtain
\begin{align}
\begin{pmatrix}
\begin{array}{c|c}
U_{i} & L_{i}^{-1} P_{i}^{T} \vec{b}_{i} \\
0 & \vec{b}_{i*}
\end{array} 
\end{pmatrix},
\label{eq:cluster_plu}
\end{align}
where $\vec{b}_{i}$ is the projection of $\vec{b}$ onto the detectors/row coordinates that are enclosed  by $C_i$. Similarly, $\vec{b}_{i*}$ is the projection onto detector coordinates not enclosed by $C_i$.
Importantly, applying the operators $L_i^{-1}$ and $P_i^{T}$ does not affect the support of $\vec{b}_i$ and $\vec{b}_{i*}$, as both these operators act solely on the support of $C_i$. 
Combining this with \cref{eq:cluster_plu}, we note that to complete the PLU factorization of $H_{[C_i \cup \{v_j\}]}$ only $\vec{b}_{i*}$ has to be reduced, which has a computational cost proportional to its weight -- crucially only a small constant for bounded LDPC matrices $H$.

We now continue by describing the collision case, where the addition of a fault node to a cluster results in the merging of two clusters.
The generalization to the merging of more than two clusters is straightforward.

Suppose that the selected fault node $v_j$ by which the cluster $C_i$ is grown is connected to a check node in the boundary $\beta(\Cj)$, with $C_i \neq \Cj$.
Let $\vec{b}$ be the column of $H$ associated with the fault node $v_j$.
Re-ordering its coordinates if necessary, we can write $\vec{b}$ as $(\vec{b}_i, \vec{b_\Cjindex}, \vec{b_*})$ where $\vec{b}_i, \vec{b_\Cjindex}$, and $\vec{b_*}$ are the projections of $\vec{b}$ on the row coordinates contained in $C_i$, $\Cj$, and neither of them, respectively.
Thus, using a block matrix notation, for the combined cluster $C_i \cup \Cj \cup \{v_j\}$, we have
\begin{align}
	H_{[C_i \cup\Cj \cup \{v_j\}]} = \begin{pmatrix}\begin{array}{c|c|c}
		H_{[C_i]} & 0      & \vec{b}_i \\ \hline
		0   & H_{[\Cj]} & \vec{b}_{\Cjindex} \\ \hline
		0   & 0      & \vec{b}_{*}
	\end{array} \end{pmatrix} .
\end{align}
By applying the PLU factorization of $H_{[C_i]}$ and $H_{[\Cj]}$ block wise, we can put $H_{[C_i \cup \Cj \cup \{v_j\}]}$ into the form
\begin{align}
\begin{pmatrix}
\begin{array}{c|c|c}
U_{i}  & 0         & L_{i}^{-1} P_{i}^{T} \vec{b}_i \\ \hline
0      & U_{\Cjindex}   & L_{\Cjindex}^{-1} P_{\Cjindex}^{T} \vec{b}_{\Cjindex} \\ \hline
0      & 0         & \vec{b}_{*}
\end{array}
\end{pmatrix}.
\end{align}
Since $U_{i}$ and $U_{{\Cjindex}}$ are, in general, not full rank, they may contain some zero rows.
As a result, the first $\lvert C_i\rvert + \lvert \Cj \rvert$ columns are not necessarily in reduced form.
To make this issue clearer, we introduce the notation $\vec{u}_{m} = L_{m}^{-1} P_{m} \vec{b}_m$ for $m \in \{ i, \Cjindex \}$ and express the above matrix as
\begin{align}
    \begin{pmatrix}
    \begin{array}{c|c|c}
    U_{i}  & 0         & \vec{u}_i \\ \hline
    0      & U_{\Cjindex}   & \vec{u}_{\Cjindex} \\ \hline
    0      & 0         & \vec{b}_{*}
    \end{array}
    \end{pmatrix} =  
    \left( 
    	\begin{array}{c|c|c}
    		U_{i, \yesrows}   & 0     & \vec{u}_{i, \yesrows} \\ \hline
    		0                  & 0     & \vec{u}_{i, \norows} \\ \hline
    		0      & U_{\Cjindex, \yesrows} & \vec{u}_{\Cjindex, \yesrows} \\ \hline
    		0      & 0                 & \vec{u}_{\Cjindex, \norows} \\ \hline
    		0      & 0         & \vec{b}_{*}
    	\end{array} \right).
     \label{eq:set_of_rows}
\end{align}
Here, by slight misuse of notation, we group the non-zero rows of $U_{m}$ in the index set $(m, \yesrows)$, and its zero rows in the set $(m, \norows)$; we regroup the coordinates of vector $\vec{u}$ accordingly. We remark that the row sets $\yesrows$ and $\bot$ are distinct from the row set $*$ identified when writing $\vec{b}$ as the combination of its projection onto row coordinates enclosed by $C_i$ and outside it. 
By identifying the appropriate row sets for the clusters $C_i$, $\Cj$ and the fault node $\{v_j\}$ as detailed in \cref{eq:set_of_rows}, we can construct a block-swap matrix 
to bring $H_{[C_i \cup \Cj \cup \{v_j\}]}$ into the form
\begin{align}
 \begin{pmatrix}
    	\begin{array}{c|c|c}
		U_{i, \yesrows}   & 0     & \vec{u}_{i, \yesrows} \\ \hline
		0      & U_{\ell, \yesrows} & \vec{u}_{\ell, \yesrows} \\ \hline
		0                  & 0     & \vec{u}_{i, \norows} \\ \hline
		0      & 0                 & \vec{u}_{\Cjindex, \norows} \\ \hline
		0      & 0         & \vec{b}_{*}
	\end{array}         
    \end{pmatrix},
\label{eq:good_form_h_combined}
\end{align}
and similarly for its PLU factors.
Since the matrix in \cref{eq:good_form_h_combined} has the same form as the one in \cref{eq:one_added_column}, the algorithm can proceed from this point onward as in the case of the addition of a single fault node to a cluster. 
In conclusion, via a swap transformation, we can effectively reduce the problem of merging two clusters to the problem of adding a single fault node to one cluster. 

\subsection{Numerical decoding simulations} \label{ssec:noise_model}
For all numerical simulations in this work, we employ a circuit-level noise model that is characterized by a single parameter $p$, the physical error probability.
Typically, the standard noise model for each time step is then to assume the following.
\begin{itemize}
    \item Idle qubits are subject to depolarizing errors with probability $p$.
    \item Pairs of qubits acted on by two-qubit gates such as CNOT are subject to two-qubit depolarizing errors \emph{after} the gate, that is, any of the 15 non-trivial Pauli operators occurs with probability $p/15$.
    \item Qubits initialized in $\ket{0} (\ket{+}$) are flipped to $\ket{1}$ ($\ket{-}$) with probability $p$.
    \item The measurement result of an $X$/$Z$ basis measurement is flipped with probability $p$.
\end{itemize}

For surface code simulations, we use the syndrome extraction circuits and noise model provided by \texttt{Stim}~\cite{gidney_stim_2021}.
We note that this noise model is similar to the one described above, however, it differs in small details such as that it combines measurement and initialization errors, ignores idling errors and applies a depolarizing channel to data qubits prior to each syndrome measurement cycle.
We perform a memory experiment for a single check side ($Z$-checks), called \texttt{surface\_code:rotated\_memory\_z} experiment in \texttt{Stim}, over $d$ syndrome extraction cycles for code instances with distance $d$.

The syndrome extraction circuits for the family of HGP codes presented in~\appref{sssec:hgp_sims} and results presented in \secref{ssec:numerical_results} are obtained from the minimum edge coloration of the Tanner graphs associated to the respective parity check matrix, see 
Ref.~\cite{xu_constant-overhead_2024} for details.
In particular, we generate associated \texttt{Stim} files of $r=12$ noisy syndrome extractions using a publicly available implementation of the aforementioned coloration circuit by Pattison~\cite{chris_github}.
In this case, the standard circuit-level noise model described at the beginning of this section is employed. 
We decode $X$ and $Z$ detectors separately using a $(3, 1)-$overlapping window decoder.
That is, for each decoding round, the decoder obtains the detection events for $w=3$ syndrome extraction cycles and computes a correction for the entire window.
However, it only applies the correction for a single $(c=1)$ syndrome extraction cycle, specifically the one that occurred the furthest in the past. For more details on circuit-level overlapping window decoding, see Ref.~\cite{scruby_high-threshold_2024}.
We have chosen $w=3$ as this was the value used in Ref.~\cite{xu_constant-overhead_2024}.
Note that it is possible that (small) decoding improvements could be observed by considering larger values of $(w,c)$ for the overlapping window decoder.

The BB codes are simulated using the syndrome extraction circuits specified in Ref.~\cite{bravyi_high-threshold_2024}, and the \texttt{Stim} files are generated using the code in Ref.~\cite{gong_toward_2024}.
There, the authors recreate the circuit-level noise model described in Ref.~\cite{bravyi_high-threshold_2024} which, up to minor details, implements the noise model described at the beginning of this section.
Similar to the HGP codes mentioned above, we decode $X$ and $Z$ decoders separately.
Analogous to our surface code experiments, we simulate for a distance $d$ code $d$ rounds of syndrome extraction and decode the full syndrome history at once.
As the BB codes are CSS codes, we decode $X$ and $Z$ detectors separately.

If not specified otherwise, we have used the min-sum algorithm for BP, allowing for a maximum of 30 iterations with a scaling factor of $\alpha = 0.625$, using the parallel update schedule.
We have not optimized these parameters and believe that an improved decoding performance, in terms of speed and (or) accuracy, can be achieved by further tweaking these parameters.

\subsection{Parallel algorithm}
We propose a parallel version of the LSD algorithm (P-LSD) in~\appref{sec:plsd} that uses a parallel data structure, inspired by Refs.~\cite{simsiri_work-efficient_2018, delfosse_almost-linear_2021}, to minimize synchronization bottlenecks. 
We discuss the parallel algorithm in more detail in~\appref{sec:plsd}. 
There, we derive a bound on the parallel \emph{depth} of P-LSD, that is, roughly the maximum overhead per parallel resource of the algorithm. 
We show that the depth is $O(\polylog(n)+\kappa^3)$ in the worst-case, where $n$ is the number of vertices of the decoding graph and $\kappa$ is the maximum cluster size.
A crucial factor in the runtime overhead of P-LSD is given by the merge and factorization operations. 
We contain this overhead by (i) using the parallel union-find data structure of Ref.~\cite{simsiri_work-efficient_2018} for cluster tracking and (ii) using parallel on-the-fly elimination to factorize the associate matrices. 
If we assume sufficient parallel resources, the overall runtime of parallel LSD is dominated by the complexity of computing the decoding solution for the largest cluster. 
To estimate the expected overhead induced by the cluster sizes concretely, we (i) investigate analytical bounds and (ii) conduct numerical experiments to analyze the statistical distribution of clusters for several code families, see~\appref{app:qldpc_clusters}.

\section{Data availability}
All simulation data is available on Zenodo~\cite{hillmann_simulation_2024}.

\section{Code availability}
The proposed algorithm and scripts to run the numerical experiments to generate the results presented above is publicly available on Github~\cite{ldpc_main}.

\section{Acknowledgments}
The authors would like to thank C.A.\@ Pattison for providing the code to generate the coloration circuits for the HGP code family via Github.

T.H.\@ acknowledges the financial support from the Chalmers Excellence Initiative Nano and the Knut and Alice Wallenberg Foundation through the Wallenberg Centre for Quantum Technology (WACQT).

This work was done in part while L.B.\@ was visiting the Simons Institute for the Theory of Computing.

L.B.\ and R.W.\ acknowledge funding from the European Research Council (ERC) under the European Union’s Horizon 2020 research and innovation program (grant agreement No.\ 101001318) and Millenion, grant agreement No.\ 101114305). 
This work was part of the Munich Quantum Valley, which is supported by the Bavarian state government with funds from the Hightech Agenda Bayern Plus, and has been supported by the BMWK on the basis of a decision by the German Bundestag through project QuaST, as well as by 
the BMK, BMDW, and the State of Upper Austria in the 
frame of the COMET program (managed by the FFG).

J. R.\@ is funded by an EPSRC Quantum Career Acceleration Fellowship (grant code: UKRI1224). J. R.\@ further acknowledges support from EPSRC grants EP/T001062/1 and EP/X026167/1.

The Berlin team has been funded by BMBF (RealistiQ, QSolid), the DFG (CRC 183), the Munich Quantum Valley, the Einstein Research Unit on Quantum Devices, the Quantum Flagship (PasQuans2, Millenion), and the European Research Council (ERC DebuQC). For Millenion and the Munich Quantum Valley, this work is the result of joint-node collaboration.

\section{Competing interests}
The authors declare no competing interests. 

\section{Author contributions}
T.H., L.B., and J.R.\@ implemented and conceived the LSD algorithm. 
T.H.\@ performed the numerical decoding simulations and implemented the overlapping window decoder. A.Q.\@ performed the numerical cluster simulations and formulated the cluster bounds.
All authors drafted the manuscript
and contributed to analytical considerations.

\bibliographystyle{quantum}
\bibliography{references}

\appendix
\section{Clusters on decoding graphs}\label{app:qldpc_clusters}
The primary motivation behind the LSD algorithm is to divide the decoding problem into smaller, more manageable, sub-problems. 
This section investigates the conditions under which such a division is feasible. 
We study the structure of clusters that form in decoding graphs of QLDPC codes under an independent and identically distributed noise model, and we relate this structure to the runtime of the parallel LSD algorithm.

As discussed in the main text, a check matrix $H$ -- whether derived from a stabilizer code or its syndrome extraction circuit -- can be interpreted as the incidence matrix of a bipartite graph, $\mathcal{G}(H) = (V_D \cup V_F, E)$, which we refer to as the Tanner graph of $H$. 
Although the Tanner graph encapsulates all the information necessary to encode the decoding problem, it is beneficial to consider the projection of the Tanner graph onto the fault nodes $V_F$ when investigating the statistical distribution of errors under a determined noise model.
We call this projected graph, with vertices $V_F$, the \emph{fault graph}, and denote it as $\mathcal{F}(H)$. 
Two vertices $v, v' \in V_F$ are connected by an edge in $\mathcal{F}(H)$ if and only if $v$ and $v'$ have a common neighbor in $\mathcal{G}(H)$, i.e., 
if and only if there exists $d \in V_D$ such that $(d, v)$ and $(d, v')$ are edges in $\mathcal{G}(H)$.

To model an error distribution on the fault graph $\mathcal{F}(H)$, we assume that each fault node is occupied randomly with probability $p$ and empty with probability $1-p$. 
Hence, for $n$ faults, we expect $np$ faults to be occupied, and $n (1-p)$ to be empty. 
This setup defines a site percolation problem, which involves studying the distribution of groups of neighboring occupied faults, called \emph{clusters}. 
We call any instance of the site percolation problem an \emph{error} defined on the fault graph $\mathcal{F}(H)$, and we expect clusters to have varying sizes and shapes depending on both the fault graph connectivity and the flip probability $p$. 

Crucially, for low enough error probability $p$, random errors are likely composed of small disjoint clusters on the fault graphs of bounded LDPC matrices $H$, and hence yield disjoint and independent decoding problems that can be solved locally~\cite{kovalev_quantum_2013, delfosse_upper_2013}. 
This argument ultimately serves as the proof that $(r, c)$-bounded LDPC codes, with distance scaling as $\sim n^{\alpha}$ for $\alpha > 0$, have a threshold \cite{kovalev_quantum_2013, gottesman_fault-tolerant_2014, aliferis_accuracy_2008}. 
In this section, we use similar statistical tools to estimate the expected performance of the LSD decoding algorithm.

Knowledge of the cluster structure of an error would enable an exact and optimal factorization of the decoding problem into smaller sub-problems. 
LSD initializes one cluster for each activated detector node on the Tanner graph and uses prior knowledge of the noise model and the fault graph structure to grow each cluster until it is valid, i.e., until it defines a solvable decoding problem.
Provided that fully grown clusters cover the original error clusters, the computational cost of a parallel implementation of LSD is bounded by the computational cost of solving the decoding problem over the biggest cluster found.

In the best scenario possible, the guided cluster growth is perfect, meaning that the clusters grown by LSD match exactly the actual error clusters. 
In this scenario, the running time of serial LSD is proportional to
\begin{enumerate}
	\item the number $\nu$ of clusters of the error and
	\item the size $\kappa$ of the largest cluster of the error.
\end{enumerate}
More broadly, the expected performance depends on the average size of the clusters of the error.

In general, if we assume sufficient parallel resources, the LSD algorithm can use $\nu$ cores and run in $O(\kappa^3)$ time in the worst case, where $\kappa$ represents the size of the largest cluster in a given instance. 
However, from a practical perspective, the largest cluster size for each sampled error (shot) is not the sole relevant figure of merit. 
As our statistical analysis suggests and our numerical investigation confirms, most errors exhibit minimal variance in their cluster size distributions, meaning that the clusters have the same size up to small fluctuations. 
Consequently, a useful practical proxy for the expected parallel runtime of LSD is the average cluster size per shot, $\kappa_{\alpha}$, rather than the worst-case cluster size.

Following this reasoning, we collect statistical data on these three quantities $\nu, \kappa$ and $\kappa_{\alpha}$, for a given noise model of interest. 
We study the expectation value of $\nu, \kappa$ and $\kappa_{\alpha}$ for: 
\begin{enumerate}[(i)]
	\item\label{point:original} errors sampled according to the noise distribution on the fault graph;
    \item\label{point:LSD} final clusters found by BP+LSD on termination of the algorithm that is, when all clusters are valid.
\end{enumerate}
The optimal LSD implementation is such that the statistics found at point (\ref{point:LSD}) match the one found at point (\ref{point:original}). 

In \figref{fig:cluster_size_stats},
we report the statistics for the fault graph obtained from the circuit-level noise simulation of the $\llbracket 144, 12, 12 \rrbracket$ code~\cite{bravyi_high-threshold_2024}. 
As we can see, for noise values below the threshold, there is no statistical difference between the two different cluster distributions, indicating that the clusters found by BP+LSD are of minimal size and hence (close to) optimal. 

\subsection{Analytic cluster bounds}
\begin{figure}[!t]
    \centering
    \includegraphics[width=1.\columnwidth]{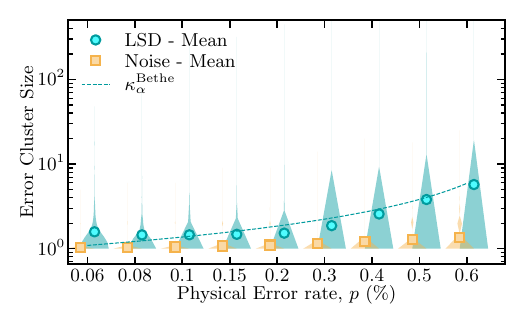}
    \caption{Distribution of the cluster size statistics of the $\llbracket 144, 12, 12 \rrbracket$  bivariate bicycle code of Ref.~\cite{bravyi_high-threshold_2024} under circuit-level noise with strength $p$. 
    Markers show the mean of the distribution while shapes are violin plots of the distribution obtained from $10^5$ samples. Yellow distributions show statistics for the optimal factorization while the blue distributions show statistics for the factorization returned by BP+LSD. The dashed line represents the expected average cluster size for the Bethe lattice, $\kappa_{\alpha}$, for $\theta = 139$.
    Markers and distributions are slightly offset from the actual error rate to increase readability.}
    \label{fig:app_clusters}
\end{figure}
In the remainder of this section, we discuss some known analytical site percolation results on the cluster distribution on regular graphs. 
Instead of investigating average clusters properties via statistical tools as proposed above, (see, e.g., \figref{fig:cluster_size_stats}), we present some upper bounds,  which mainly make use of approximation of regular graphs as regular trees~\cite{stauffer_introduction_2017,Grimmett}. 

If $H$ is an LDPC matrix with constant row and column weight $r$ and $c$, respectively, the associated fault graph $\mathcal{F(H)}$ is a regular graph of degree $\theta = c(r-1)$, with ${\theta \cdot n}/{2}$ edges.
Let $n_s^p(v)$ be the number of clusters of size $s$ containing the vertex $v$. 
In general, this takes the form
\begin{align}
    p^s (1-p)^{|\mathrm{perimeter}|},
\end{align}
where the perimeter of a cluster is the set of non-occupied nodes adjacent to at least one node in the cluster. 
The probability that the node $v$ belongs to a particular cluster of size $s$ is $n_s^p(v)s$, as $v$ can be any of the $s$ nodes in the clusters.  
On our percolation problem, by assumption, the probability 
that a given node is occupied is $p$. 
Moreover, if a node is occupied, it belongs to a cluster of size $\tilde s$ for some $\tilde s = 1, \dots, n$, where $n$ is the number of nodes in the graph. 
Hence, for a given node in the graph, the probability $p$ of being occupied must equate the sum of the probabilities of belonging to a cluster of size $\tilde s=1, \dots, n$. 
Hence, for an arbitrary node in the graph, it holds that
\begin{align}
\label{eq:prob_a_cluster}
    \sum_{s} n_{s}^p(v) s = p.
\end{align}
That is, the probability $n_{s}^p(v)$ that a node belongs to an $s$-cluster times the number of nodes/choices for that cluster.
A quantity of interest for decoding is %
an estimate of the size of the cluster a random node belongs to. In other words,  the probability $w_s$ that a node is occupied and belongs to an $s$-cluster.
Explicitly, using~\Cref{eq:prob_a_cluster}, we find
\begin{align}
    w_s = \frac{n_s^p(v)s}{p} = \frac{n_s^p( v)s}{  \sum_{\tilde s} n_{\tilde s}^p(v) \tilde s},
\end{align}
and therefore the average cluster size is given by
\begin{align}
\label{eq:avg_clust}
    \kappa_{\alpha} &= \sum_{s} w_{s} s \nonumber\\
    &= \frac{1}{p}\sum_{s} n_s^p(v) s^2.
\end{align}
For a general graph of bounded degree $\theta$, these quantities are difficult to estimate. 
The distribution of cluster sizes takes asymptotically the form $s^{-\tau} e^{-s/s_\xi}$ for large sizes $s$ \cite{Grimmett}, with some characteristic cluster size $s_\xi$ and for a suitable $\tau>0$, but again, for our purposes, they are not
easy to assess.

Nonetheless, we can find some upper bounds by looking at the same percolation problems on the Bethe lattice. The Bethe lattice is an infinite tree where all vertices have the same degree. 
For instance, 
on each $\theta$-bounded graph, the size of the perimeter of a $s$-cluster is upper-bounded by the size of the perimeter of an $s$-cluster on the Bethe lattice of degree $\theta$, that is the same for all $s$-clusters and it is maximal across the perimeters of $s$-clusters for $\theta$-bounded graphs,
taking the value $(\theta - 2)s + 2$.
 
For the Bethe lattice, we can use self-similarity to compute the average cluster size $\kappa_{\alpha}^{\mathrm{Bethe}}$. 
If $v$ is occupied, we can think of the cluster it belongs to as made up of the node $v$ itself plus the sub-clusters i.e., subsets of connected vertices in that cluster, rooted in its occupied neighbors. 
Each occupied neighbor contributes to the cluster rooted in $v$ with itself (add 1) and the sub-clusters not rooted in $v$ itself, which at most $\theta-1$; by self-similarity, these sub-clusters have same average size as the cluster rooted in $v$, and contribute to the count adding at most $(\theta-1)\kappa_{\alpha}^{\mathrm{Bethe}}$ vertices. 
Thus, if the vertex $v$ is occupied, it belongs to a cluster of size $1 + (\theta -1)\kappa_{\alpha}^{\mathrm{Bethe}}$. 
As each neighbor of the vertex $v$ is empty with probability $(1-p)$, and occupied with probability $p$, we find
\begin{align}
    \kappa_{\alpha}^{\mathrm{Bethe}} &= p\left(1+ (\theta-1)\kappa_{\alpha}^{\mathrm{Bethe}}\right),
\end{align}
and via substitution:
\begin{align} 
    \kappa_{\alpha}^{\text{Bethe}} = \frac{p}{1-p(\theta-1)}.\label{eq:cluster-size-bethe}
\end{align}
In conclusion, provided that the decoding problem of our interest is defined on a parity check matrix of bounded column and row weight, we can upper bound the expected average cluster size with the expected average cluster size $\kappa_{\alpha}^{\mathrm{Bethe}}$ on the Bethe lattice~\cite{stauffer_introduction_2017,Grimmett}.

As an example of this upper bound via the Bethe lattice, in \cref{fig:app_clusters}, we plot cluster size distributions against the physical error rate $p$ in a circuit-level noise model for the $\llbracket 144, 12, 12\rrbracket$ code. 
The yellow distributions show the distribution of cluster sizes of the optimal factorization, and the blue distributions are obtained from a BP-guided implementation of an LSD decoder (BP+LSD), where the samples are post-selected on errors for which BP does not converge.
That is, errors correctable by BP only are not included here for the BP+LSD distributions.
The dashed line represents the average cluster size for the Bethe lattice as in \cref{eq:cluster-size-bethe} with the parameter $\theta = 139 $.

We expect the LSD cluster distribution to match the cluster distribution on a fault graph with a higher vertex degree than the original fault graph, due to the uncertainty in the growth step guided by the error likelihood vector $\lambda$. 
For the fault graph of the $\llbracket 144, 12, 12\rrbracket$ code considered here, which has average vertex degree $\theta_{\alpha} = 103$, and maximum vertex degree $\theta_{\mathrm{max}} = 166$, the fitting parameter $\theta = 139$ supports our heuristic argument. 
We note that for lower noise rates, specifically when $p < 0.1 \%$, the Bethe lattice's average cluster size distribution does not upper bound the LSD distribution. 
This is consistent with the fact that LSD is never called when the likelihood of error vector $\vec{\lambda}$ perfectly describes the actual error, i.e., when BP converges. 

In conclusion, for low noise rates, the clusters found by LSD are small and closely approximate the optimal size of the clusters induced by the original error.
In the low-error regime, the local decoding problem within each cluster is efficiently solvable.

\section{Parallel implementation}\label{sec:plsd}
In this section, we propose a parallel version of LSD and analyze its time complexity.
The central observations are that clusters can be grown and solved (using the on-the-fly elimination technique outlined in Methods 4.2) in parallel and hence potential synchronization issues arise only due to cluster merges.

To enable efficient parallel merging of clusters, we represent the clusters by a parallel version of the union-find data structure as proposed in
Ref.~\cite{simsiri_work-efficient_2018}, where the authors prove that the union-find operations can be performed in parallel polylogarithmic time on each parallel resource with such a data structure.
We detail the key steps of the parallel implementation of the LSD algorithm, P-LSD, in~\Cref{alg:plsd}.

\begin{enumerate}
\item (Parallel) initialization: we create a cluster $C_i$ for each flipped detector node $s_i$ in the syndrome. 
We then compute the set of candidate fault nodes at the boundary of each cluster: $\Lambda(C_i) = \Gamma(\{s_i\})$ --  see Def 4.3.
Clusters are managed as a \emph{union-find} forest in the parallel data structure $\mathbb{U}$, see 
Ref.~\cite{tarjan1975efficiency}.

\item (Parallel) pre-growth step: for each invalid cluster in $\mathbb{U}$, the \textsc{Find\_candidate} function is used to identify, among its candidate fault nodes, the one with the maximum probability to be in error according to the soft information vector $\vec{\lambda}$. 
We call the set of all these selected nodes growth\_faults, one for each invalid cluster. 

\item (Parallel + synchronization) growth step: add one new fault node to each cluster and merge colliding clusters [Def. 4.4 (\emph{Cluster collisions})] using the parallel union-find data structure $\mathbb{U}$ and its efficient \textsc{union-find} operations.
The \textsc{parallel\_union} routine~\cite{simsiri_work-efficient_2018} is used to efficiently determine which set in $\mathbb{U}$ a node belongs to and to efficiently merge any two sets in $\mathbb{U}$ that correspond to colliding clusters.

\item (Parallel) validity check: Apply on-the-fly elimination in parallel for each cluster, see Methods 4.2..
\item Termination: Compute a local solution for each cluster and combine them to return the overall correction.
\end{enumerate}

\begin{algorithm}[!bh]
\SetAlgoLined
\DontPrintSemicolon
\SetKwBlock{DoParallel}{do in parallel}{}
\SetKwBlock{Synch}{synchronize}{}
\caption{Parallel localized statistics decoding (P-LSD) \label{alg:plsd}}
$H$: decoding matrix\; 
$\mathbf{s}$: syndrome support set\;
$\vec{\lambda}$: fault nodes error soft information\;
$\mathbb{U}$: parallel union-find data structure \;
\DoParallel({\quad for $s_i \in \mathbf{s}$}){
$\mathbb{U}$ $\leftarrow$ \textsc{initialize\_cluster}($s_i$)\;
}
\BlankLine
\Synch{}
\While{$\exists$ invalid clusters \label{line:iteration}}{
    growth\_faults $\leftarrow$ \textsc{find\_candidate}($\mathbb{U}$, $\vec{\lambda}$)\;
    \textsc{parallel\_union}($\mathbb{U}$, growth\_faults)\;
  
    \DoParallel({\quad for invalid $C_i \in \mathbb{U}$}){
        $C_i$.\textsc{on\_the\_fly\_elimination}()\;
	}
 }
\end{algorithm}

We now analyze the parallel time complexity of the main routine in P-LSD.
To this end, we derive a bound on the parallel \emph{depth} of the main routine in P-LSD, where the parallel depth is defined as the maximum number of dependent sequential steps in the computation.

\begin{enumerate}
    \item Initialization: $O(1)$ parallel depth.
    This includes the identification of the candidate fault nodes set for each cluster, which has size $O(1)$ for bounded LDPC matrices $H$. 
    \item Pre-growth step: iterate over the candidate fault nodes for each cluster, and identify the most suitable fault node given the soft information vector $\vec{\lambda}$.
    This step can be performed in logarithmic complexity in the size of the boundary using, e.g., a Fibonacci heap. 
    Maintaining a list of candidate fault nodes [Def. 4.3 (\emph{Cluster-boundary}] for each cluster throughout the algorithm incurs only a constant depth overhead.
    \item Growth-step: using the methods presented in~\cite{simsiri_work-efficient_2018}, growing and merging of clusters can be done in  $O(\polylog(n))$ depth, where $n$ is the total number nodes in the decoding graph.
    \item Validity check: the depth is dominated by the cost of the PLU factorization of each cluster matrix. 
    That is $O(|C_i|^3)$, where $|C_i|$ indicates the number of columns of the cluster $H_{[C_i]}$.
    \item Termination: finding a local solution given the PLU factorization of the check matrix of a cluster has cost $O(|C_i|^2)$.
\end{enumerate}

In conclusion, the overall parallel depth for an iteration of the while loop in \cref{line:iteration} of \cref{alg:plsd} is in $O(\polylog(n) + \kappa^3)$, where $\kappa$ is the maximum cluster size.
Crucially, the maximum cluster size $\kappa$ is expected to be small for QLDPC codes and low enough error rates, see  \cref{app:qldpc_clusters}. 
In fact, the average cluster size $\kappa_\alpha$ can be bounded for regular graphs, such as the Bethe lattice by $\kappa_\alpha \leq \frac{1+p}{1-(\theta-1)p}$, where $\theta$ is the vertex degree of the graph and $p$ is the physical noise strength.

The $O(\polylog(n) + \kappa^3)$ depth bound is a loose upper bound and in practice the runtime of the algorithm depends on multiple additional factors such as the concrete implementation, overhead of parallelization, and the number of parallel resources available. 
The overall \emph{work}, i.e., the total number of steps the algorithm performs across all processes, depends on the number of growth steps.
In the worst case, the number of growth steps is proportional to the number of edges in the decoding graph, and the parallel union-find data structure will have overall work almost linear in $n$, more precisely  $O\left(n \alpha(n)\right)$, where $\alpha(\cdot)$ is the inverse Ackerman function, which is $\alpha(n)\leq 3$ for any practical situation~\cite{delfosse_almost-linear_2021, simsiri_work-efficient_2018, tarjan_worst_1984}.
However, if almost $n$ growth steps were necessary for P-LSD to terminate, the computed error estimate would probably have weight above the percolation threshold of the graph, in which case decoding introduces a non-trivial logical error with high probability.
In other words, we expect almost $n$ growth steps only in instances of the decoding problem that have an intrinsically non-local structure, meaning instances that can not be divided into smaller decoding problems. 
In such a case, there is no benefit in parallelization and hence the parallel runtime complexity is only marginally meaningful.

\section{Numerical experiments \label{app:numerical_experiments}}
In this section, we provide further details on the numerical decoding experiments presented in the main body of the manuscript. 

\subsection{Surface code decoding \label{sssec:surface_codes_sims}}

The surface code~\cite{kitaev_fault-tolerant_2003, dennis_topological_2002} is a ``matchable'' code, that is, 
qubits participate in at most two X and two Z stabilizer measurements. 
Hence, syndromes come in pairs and thus the code is 
decodable using the \emph{minimum-weight perfect matching} (MWPM) algorithm~\cite{higgott_pymatching_2021, higgott_sparse_2023}.
The surface code is one of the best-known quantum error correcting codes and thus can be seen as a crucial ``benchmark'' case for decoders. 
This aspect is facilitated by the high-quality open-source implementation of~\texttt{Stim}~\cite{gidney_stim_2021}---a tool that can be used for automated circuit noise simulations, which also makes comparing numerical evaluations between different authors consistent.
We refer the reader to Refs.~\cite{gidney_stim_2021, higgott_sparse_2023} for more details on \texttt{Stim}.

The BP+UF algorithm~\cite{higgott_improved_2023} is a combination of BP and Union-Find, where the BP soft information is used to guide the cluster growth for the Union-Find algorithm. 
Note that in the case of a matchable code, the validity check for clusters, which is equivalent to solving systems of linear equations in general, can be replaced by a simple parity computation of the support vector of enclosed, flipped detector nodes and hence is more efficient. 
Moreover, computing a solution for the Union-Find clusters can be done using a simple algorithm based on spanning-tree construction and thus is also efficient for matchable codes.
Note that BP+UF and BP+LSD share an identical cluster growth strategy and therefore should converge to the same set of valid clusters, with the difference that BP+UF applies the peeling decoder~\cite{delfosse_almost-linear_2021} for finding a valid correction for each cluster, while the LSD algorithm performs a (partial) inversion of the cluster check matrices $H_{[C]}$ to determine the correction.x

\subsection{Hypergraph product codes \label{sssec:hgp_sims}}
We investigate the decoding performance of a family of hypergraph product (HGP) codes~\cite{tillich_quantum_2014} with rate \mbox{$k/ n \geq 1/25$}. 
The codes are single-shot decodable~\cite{bombin_single-shot_2015, higgott_improved_2023}, that is, a constant-sized decoding window is sufficient for decoding codes of arbitrary size. 
This family of codes has been previously proposed and investigated in Ref.~\cite{grospellier_combining_2021} under a code capacity noise model, and later in Ref.~\cite{tremblay_constant-overhead_2021} under a circuit-level noise model. 
In both cases, decoding was achieved by a combination of BP and \emph{small set flip} (SSF)~\cite{leverrier_quantum_2022}.
Additionally, in Ref.~\cite{xu_constant-overhead_2024} the authors investigated this code family under a phenomenological noise model derived from a circuit-level noise model that is inspired by a potential implementation in neutral atom arrays using BP for the bulk of the decoding problem and BP+OSD in the last decoding round to ensure a correction that projects state back into the code space.
This family of codes is obtained from random $(3, 4)-$regular Tanner graphs with girth at least 6.
Since the construction is random, following Ref.~\cite{tremblay_constant-overhead_2021}, we generate 100 instances of each check matrix and select the best one after performing code capacity simulations.
Interestingly, we find that in some cases the performance difference between the best and the worst performing codes is close to a factor of 10.
Here, we investigate the sub-threshold performance for a circuit-level noise model for 4 instances of the family using our proposed BP+LSD decoder.

\subsection{Bivariate bicycle codes \label{ssec:bb_codes_sims}}
We conclude our numerical decoding benchmarks by decoding instances of \emph{bivariate bicycle} (BB) codes recently investigated in Ref.~\cite{bravyi_high-threshold_2024}, and originally proposed in Ref.~\cite{kovalev_quantum_2013}.
The circuit-level noise simulations use the publicly available implementation by Ref.~\cite{gong_toward_2024} of the highly optimized syndrome extraction circuits described in Ref.~\cite{bravyi_high-threshold_2024}.  Detector errors are sampled using \texttt{Stim}~\cite{gidney_stim_2021}.
Similar to our surface code experiments, we simulate $d$ rounds (where $d$ is the code distance) of syndrome extraction and decode the full syndrome history at once.
The BB code family are CSS codes where the $X$ and $Z$ detectors can be decoded separately.

In \figref{fig:ibm_codes_full}, we show the logical $Z$ error rate per syndrome cycle $p_{L_{Z}}$ for the BB code family decoded with BP+LSD-0. 
Our results show that \hbox{BP+LSD-0} achieves comparable decoding performance to the BP+OSD-CS-7 decoding results presented in Ref.~\cite{bravyi_high-threshold_2024} (where BP+OSD-CS-7 refers to the ``combination sweep'' strategy for OSD higher-order processing with order $w=7$, see Ref.~\cite{roffe_decoding_2020}). 
This implies that, at least for the case of the BB code family, higher-order reprocessing has minimal impact when decoding in the sub-threshold regime. 
See \Cref{app:higher_order_lsd} for further discussion of higher-order reprocessing.

Another interesting observation from our numerical decoding simulations of BB codes is that it is sufficient to run BP+LSD with a low number of BP iterations: the BP+LSD simulations shown in \figref{fig:ibm_codes_full} of the main text were run with maximum $30$ BP iterations, compared to the maximum number of iterations of $10,000$ for the BP+OSD-CS-7 simulations in Ref.~\cite{bravyi_high-threshold_2024}.
For BP-OSD, it is preferable to run BP until convergence to avoid the cost of computing the inverse of the detector check matrix. 
In OSD, matrix inversion is performed globally with worst-case cubic runtime complexity in the number of fault nodes in the Tanner graph $n$. 
Therefore, the OSD runtime does not scale with the error rate and is independent of the number of flipped detectors.
In contrast, this is not the case for our proposed BP+LSD decoder.
When run in parallel, the LSD algorithm has an expected runtime in $O(\kappa^3)$, where $\kappa$ is the maximum cluster size. 
Further to this, the number of clusters upon initialization is identical to the number of invalid detectors. We therefore expect the decoding time to be proportional to the physical error rate. As such, in the sub-threshold regime, LSD post-processing is much less costly than BP+OSD. As a result, it is not detrimental to the decoder's runtime if a large portion of the computational load is shouldered by the LSD post-processing.

As mentioned in the main text, our implementation of BP+LSD uses BP solely to guide cluster growth. For the codes studied in this work, it is possible to gain sufficient BP soft information after approximately $30$ BP iterations.

\subsection{Runtime estimation}\label{sec:runtime-estimation}
In this section, we conduct preliminary simulations to investigate the runtime of the proposed prototypical open-source implementation of the decoder available at~\cite{ldpc-package}.
We would like to highlight that we leave an in-depth investigation of a fully parallel and optimized implementation as future work.
The results presented here should be seen as a mere representation of the runtime of the currently available software implementation acting as a benchmark for future developments, that is, in particular, it is not meant as a real-time decoder.

\begin{figure}[t]
    \centering
    \includegraphics[width=\columnwidth]{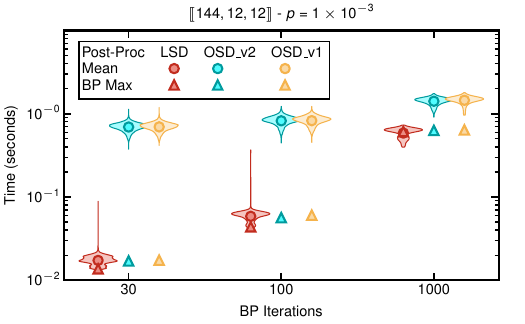}
    \caption{
    Timing statistics for circuit-level noise decoding of the $\llbracket 144,12,12 \rrbracket$ bivariate bicycle code at a physical error rate of $p=0.1\%$.
  The timing results are post-selected on BP non-convergence.
  Triangle markers indicate the longest time BP took if it converged.
    }
    \label{fig:runtime-ps}
\end{figure}

\figref{fig:runtime-ps} shows timing statistics for BP+LSD and BP+OSD.
The violin plots depict the distributions of the timings of BP+OSD and BP+LSD, where circles represent the means.
This data is post-selected on cases where LSD/OSD is called, i.e., where BP does not converge, such that one measures more accurately the performance of the post-processing routines.
To highlight the overhead introduced by the post-processor, triangles depict the maximum runtime of BP on shots for which it converged, presenting a lower bound on the achievable timing statistics.
For the BP decoder, we have used a parallel schedule and min-sum update rules.

From the plot, it is evident that for the current software implementation of BP+LSD, the LSD step only marginally increases the overall runtime, as opposed to for BP+OSD, where the OSD post-processing time induces significant runtime overhead compared to the runtime of BP alone.

\begin{figure*}
     \centering
    \includegraphics[width=0.49\textwidth]{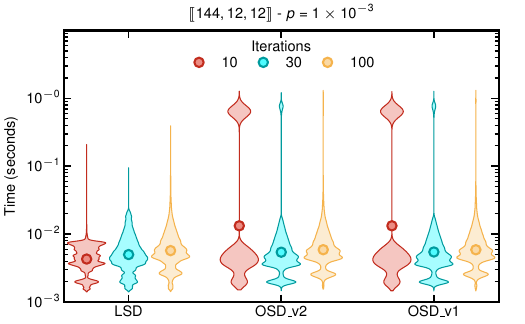}
    \includegraphics[width=0.49\textwidth]{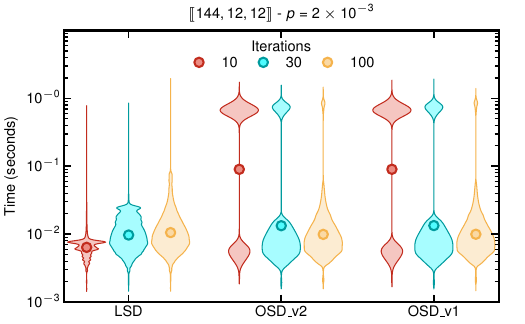}
    \includegraphics[width=0.49\textwidth]{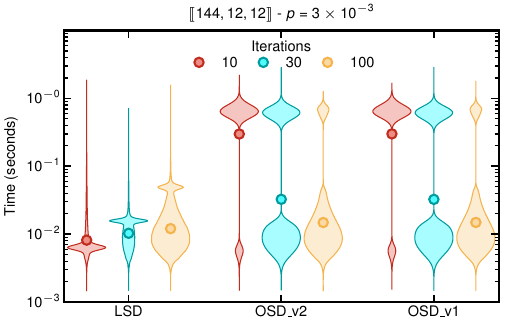}
    \includegraphics[width=0.49\textwidth]{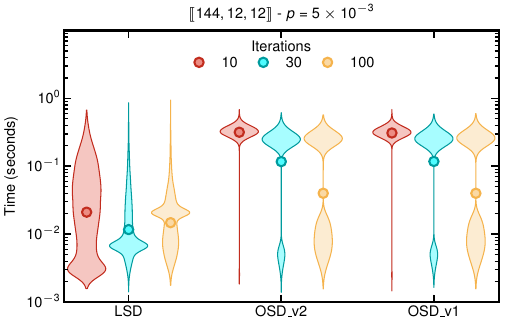}
    \includegraphics[width=0.49\textwidth]{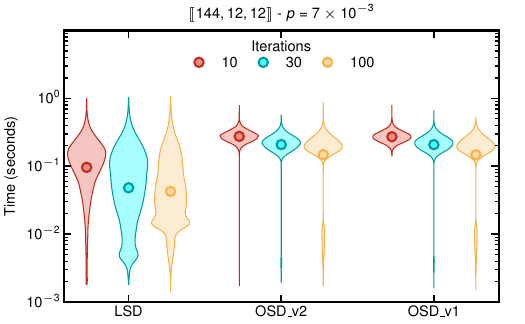}
    \hspace{0.49\textwidth}
     \caption{Runtime for the decoding of the $\llbracket 144,12,12\rrbracket$ BB code instance at physical noise rate $p=0.1\%$ (top left), $p=0.2\%$ (top right), 
     $p=0.3\%$ (center left), 
     $p=0.5\%$ (center right), 
     and $p=0.7\%$ (bottom left), that is, above the pseudo-threshold of the code~\cite{bravyi_high-threshold_2024}.
     Circle markers show average decoding time per shot. }
     \label{fig:runtime-not-ps}
\end{figure*}

To estimate the runtime behaviour of the decoder in the ``realistic scenario'', for instance, in a numerical experiment, we also investigate timing data that is not post-selected on the BP-non convergence.
\figref{fig:runtime-not-ps} depicts the total timing statistics of a BP+LSD/OSD call, regardless of BP convergence, i.e., for some samples, BP might converge and LSD/OSD is not called.
We observe that with an increased number of BP iterations, the decoding times decrease, perhaps due to more BP convergence and thus fewer post-processor calls. 
The distributions of BP+OSD exhibit a bimodal form, corresponding to the cases where BP converges or does not. For a larger number of BP iterations, this effect diminished, again perhaps due to the increased convergence rate of BP.
This is not the case for BP+LSD, whose distributions do not exhibit this bimodality form because the overhead induced by LSD is (on average) comparable to BP (and lower than OSD), but its distributions feature long tails.
\clearpage

\section{Higher-order LSD} 
\label{app:higher_order_lsd}
Standard LSD achieves comparable decoding performance in terms of logical error rate as BP+OSD.
However, improved decoding accuracy can be achieved with higher-order reprocessing, i.e., BP+OSD$-w$ for $w>0$~\cite{panteleev_degenerate_2021, roffe_decoding_2020}.
In a similar vein, we propose higher-order LSD,  LSD$-\mu$, and denote by LSD$-0$ standard LSD without further reprocessing. 

The central idea of higher-order LSD is to add an additional reprocessing step to LSD$-0$ to achieve better decoding accuracy.
Once LSD$-0$ has terminated, meaning that all clusters are valid and local solutions have been computed, we conduct, up to $\mu$ additional growth step on each valid cluster. 
On the grown clusters, we then apply standard OSD reprocessing: we do this separately on each cluster matrix and therefore we preserve LSD locality.
We note that: 
\begin{enumerate}
    \item The additional growth steps may lead to further cluster merges in general.
    \item Clusters cannot become invalid due to the additional growth steps.
    \item The number of columns added to each cluster is an important parameter that affects decoding performance.
\end{enumerate}
As such, the choice of the parameter, $\mu$, is important for the runtime as well as the error correction performance of higher-order LSD. 

Let us illustrate why it is important for higher-order reprocessing to perform additional growth steps of the clusters once the validity condition \mbox{$s_{[C_i]} \in \text{\textsc{image}}(H_{[C_i]})$} is fulfilled.
Technically speaking, this is required to ensure that the rank of the union of clusters $H_{[\cup_i C_i]}$ is close to the rank of the full matrix $H$.
Otherwise, not all possible codewords of the classical code $H$ can be represented by the clusters $H_{[C_i]}$, a condition required in higher-order OSD reprocessing.
To this end, we apply a heuristic approach where the parameter $\mu$ is chosen as a (constant) fraction of the total number of columns of the overall decoding matrix $H$. 
If $\mu$ is large enough, the cluster matrices have the desired property with high probability.
\begin{figure*}[t]
    \centering
    \includegraphics[width=1.\textwidth]{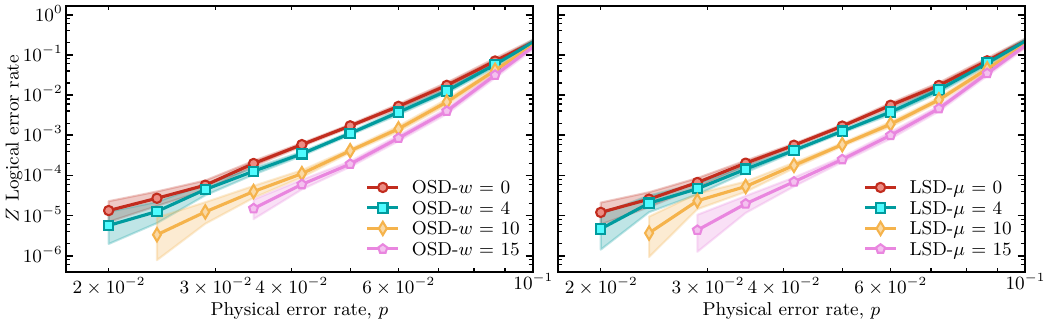}
    \caption{Effect of higher-order reprocessing routines when decoding the $\llbracket 882, 48, 16 \rrbracket$ code with (a) BP+OSD$-w$ and (b) BP+LSD$-\mu$. We observe that within the sampling variance, the logical error rates obtained from both decoders are identical.}
    \label{fig:panteleev_osd_lsd_comparison}
\end{figure*}
Our numerical findings, which we discuss in the remainder of this section, demonstrate that higher-order LSD can achieve improved decoding performance compared with LSD$-0$.
Moreover, the LSD$-\mu$ performance improvements are on par with performance improvements obtained from global OSD$-w$ reprocessing.

For the local OSD reprocessing in LSD$-\mu$, we use parameters chosen after initial decoding experiments.
We leave an in-depth exploration of LSD$-\mu$ and corresponding optimizations open for future work.

\paragraph{Comparison of BP+LSD$-\mu$ and BP+LSD$-w$}

To demonstrate the effect of LSD$-\mu$, we compare its decoding performance to OSD$-w$ for a lifted product (LP) code instance from Ref.~\cite{panteleev_degenerate_2021}, in which the authors originally proposed BP+OSD$-w$ for decoding QLDPC codes.
For demonstration purposes, we perform a code capacity experiment where we sample error vectors $\vec{x}$ from an independent and identically distributed noise model, where each qubit is flipped with probability $p$.
Using different decoders, we try to infer whether a logical $Z$ error occurred on any of the logical qubits.
For the higher-order reprocessing routine, we use an exhaustive search (OSD-E) of order $w = \mu$ for both LSD$-\mu$ and OSD$-w$.
That is, for LSD$-\mu$ we choose, heuristically, the number of additional growth steps per valid cluster to equate the local reprocessing order of the OSD-E routine.

The results depicted in~\Cref{fig:panteleev_osd_lsd_comparison} demonstrate, on the one hand, that higher-order LSD achieves a lower logical error rate for larger values of $\mu$, and on the other hand, that LSD$-\mu$ can achieve a decoding performance that matches LSD$-w$ within the sampling variance.
Note that, however, a direct comparison of the parameters $\mu$ and $w$ is not possible since they have different meanings.
Additionally, we note that performing 10 or 15 additional growth steps per cluster for the $\llbracket 882, 48, 16 \rrbracket$ code corresponds to growing each cluster additionally by around $1\%$ to $1.5\%$ of the total number of columns in the check matrix $H_X$, implying that ``microdosing'' LSD-$\mu$ is sufficient.

\begin{figure}[!b]
    \centering
    \includegraphics[width=\columnwidth]{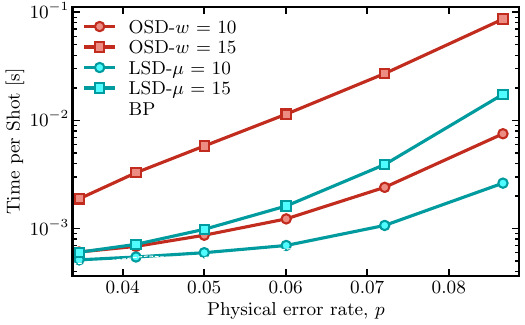}
    \caption{The time-per-shot for the $\llbracket 882, 48, 16 \rrbracket$ LP code instance for different orders of OSD and LSD reprocessing. 
    Since our timings include the BP stage of the decoder as well, we find that LSD$-\mu$ with $\mu = 10$ is limited by the serial implementation of the BP decoder and other overhead for low error rates. At larger error rates, its runtime matches the one of BP+LSD$-0$.}
    \label{fig:time_per_shot}
\end{figure}

\paragraph{Runtime comparison of higher-order global OSD and higher-order LSD.}
To estimate the runtime of our serial implementation of BP+LSD$-\mu$ and to compare it against the global BP+OSD$-w$ decoder, we plot the runtime per shot -- including the BP iterations and sampling overhead -- in seconds for various physical noise rates for the code capacity noise model described in the previous paragraph.
The results are shown in~\Cref{fig:time_per_shot}.
This figure does not aim to qualitatively 
 highlight the runtime of the decoder, but instead intends to demonstrate that even with the added higher-order reprocessing and additional growth steps, the locality of the decoding problem is, on average, preserved. In fact, if this were not the case, the average decoding time for LSD$-\mu$ would be on par with the decoding time of the global OSD$-w$ decoder.
The plot in \cref{fig:time_per_shot} indicates that, even for serial  LSD-$\mu$, there is a significant runtime improvement with respect to BP+OSD$-w$, due to the reduced size of the inversion problem, as well as the reduced search space for the OSD-E reprocessing routine on each cluster.

\begin{figure}[!t]
    \centering
    \includegraphics{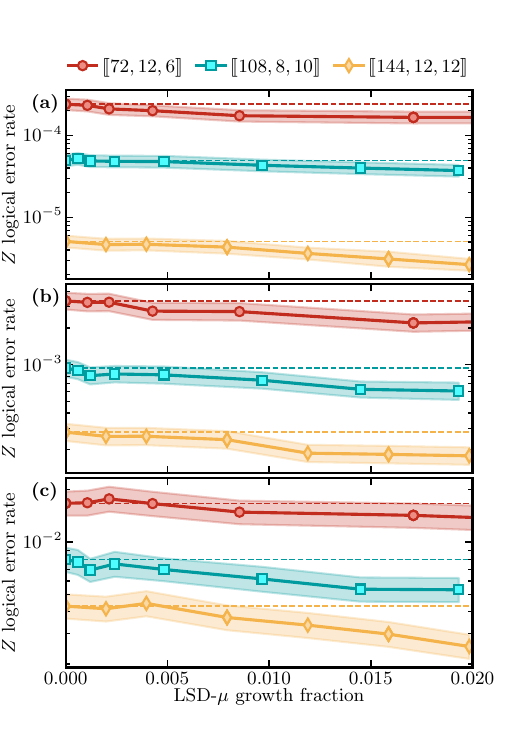}
    \caption{Performance improvements of the bivariate bicycle codes of Ref.~\cite{bravyi_high-threshold_2024} for higher-order reprocessing with the LSD decoder.
    Each cluster is grown a certain fraction $\mu$ of the total number of fault nodes in the decoding graph after an initial solution to the inversion problem is obtained. Then, a local reprocessing using the standard OSD-CS-4 method is performed.
    Each panel shows a different error rate of the circuit-level noise model. (a) $p = 0.001$, (b) $p = 0.002$, and (c) $p = 0.003$.
    The shading indicates hypotheses whose likelihoods are within a factor of 1000 of the maximum likelihood estimate, similar to a confidence interval.}
    \label{fig:lsd-mu-bb}
\end{figure}

\paragraph{Higher-order LSD decoding on bivariate bicycle codes}
Here, we investigate the decoding performance of LSD$-\mu$ for the three instances of the bivariate bicycle codes from Ref.~\cite{bravyi_high-threshold_2024} under the circuit-level noise model described in  Methods 4.3.
In~\figref{fig:lsd-mu-bb} the $x$-axis represents the additional growth steps per cluster of the LSD$-\mu$ routine as a fraction of the total number of fault nodes in the detector error model.
For each data point, we have fixed the local OSD reprocessing method used in LSD$-\mu$ to OSD-CS-4.
We observe that for all noise values, the $Z$ logical error rate is consistently reduced by increasing the number of growth steps of LSD$-\mu$.
The improvements are relatively small, which is, however, also the case when applying OSD reprocessing globally.
Therefore, we explicitly highlight the LSD-$0$ performance for each code instance and noise strength with horizontal dashed lines.
We observe that the largest code, the $\llbracket 144, 12, 12 \rrbracket$ code, benefits the most from the higher-order reprocessing.

\paragraph{Adaptive-order reprocessing.}
It is in principle possible to \emph{adaptively} choose the order of higher-order reprocessing routines based on the observed syndrome $\vec{s}$, the soft information vector $\vec{\lambda}$, and the linear dependencies encountered during the construction of the information set $I$. 
This is in contrast to current reprocessing routines, that have been implemented in the context of quantum error correction, which fix the order in advance.
In the classical error correction literature this adaptive strategy has been described by Fossorier~et~al.\ in 1998~\cite{fossorier_reliability-based_1998}.
There, the authors describe a (not necessarily efficient) \emph{covering test} that allows, for each received syndrome, to obtain statements about whether an exhaustive search reprocessing routine of order $w_2$ can improve upon an order $w_1<w_2$ reprocessing routine. 
Additionally, assuming that the soft information vector $\vec{\lambda}$ closely represents fault probabilities, this test bounds the probability that order $w_2$ will improve over order $w_1$ reprocessing.
As a result, given a certain target error rate, one can adaptively choose the reprocessing order $w$ to achieve that targeted error rate.
We leave the implementation and potential adaption of the covering test to the decoding problem of quantum error correction codes as an interesting open question for future work.

\pagebreak

\end{document}